# The Role of Mandated Mental Health Treatment in the Criminal Justice System

Rachel Nesbit[*]

November 14, 2023

### Abstract

Mental health disorders are particularly prevalent among those in the criminal justice system and may be a contributing factor in recidivism. Using North Carolina court cases from 1994 to 2009, this paper evaluates how mandated mental health treatment as a term of probation impacts the likelihood that individuals return to the criminal justice system. I use random variation in judge assignment to compare those who were required to seek weekly mental health counseling to those who were not. The main findings are that being assigned to seek mental health treatment decreases the likelihood of three-year recidivism by about 12 percentage points, or 36 percent. This effect persists over time, and is similar among various types of individuals on probation. In addition, I show that mental health treatment operates distinctly from drug addiction interventions in a multiple-treatment framework. I provide evidence that mental health treatment's longer-term effectiveness is strongest among more financially-advantaged probationers, consistent with this setting, in which the cost of mandated treatment is shouldered by offenders. Finally, conservative calculations result in a 5:1 benefit-to-cost ratio which suggests that the treatment-induced decrease in future crime would be more than sufficient to offset the costs of treatment.

**JEL codes: I12, I18, K42**

---
[*]University of Maryland, 3114 Tydings Hall, 7343 Preinkert Dr., College Park, MD 20742. Email: rnesbit@umd.edu

# 1 Intro

Poor mental health is widely prevalent and has been growing over time, with around 58 million adults in the United States suffering from a mental illness in 2021 (SAMHSA, 2022).[1] Poor mental health is also widely impactful; it has direct negative effects on physical and social health and is highly intertwined with other aspects of life. For example, it contributes to upwards of 40 percent of all illnesses under the age of 65 (Layard, 2013). Beyond (or as a consequence of) these direct negative effects on health, mental illness has been associated with behaviors like absenteeism at work, decreased educational attainment, and crime (Bubonya, Cobb-Clark, and Wooden, 2017; Burton et al., 2008; Breslau et al., 2008; Mojtabai et al., 2015).

Poor mental health is particularly prevalent in the criminal justice system. Among the approximately five million men on probation in 2009, 33 percent met the threshold for any mental illness - nearly twice the prevalence in the general population at the time (Feucht and Gfroerer, 2011).[2] Individuals with a mental illness can struggle to make rational, welfare-maximizing decisions, which may play a critical role in their interactions with the criminal justice system. Since the medical field has shown that therapy and medication can reduce symptoms of mental illness (Cronin, Forsstrom, and Papageorge, 2020; Paykel et al., 1999), it may be the case that mental health treatment can reduce behavioral outcomes such as crime. That reduction could have large benefits for both treated individuals and society. However, though meaningful correlational evidence exists, there is little direct causal evidence on the effect of therapy and psychiatric medication on outcomes such as criminal behavior, or whether those interventions could be carried out in a cost-effective way.[3]

This paper evaluates the causal impact of mandated mental health treatment on the likelihood of committing a future crime. Specifically, it focuses on how being assigned mental health treatment at the time of probation impacts recidivism over the next five years. To do this I exploit judge variation in court-mandated mental health treatment in North Carolina, using the universe of criminal court cases from 1994 to 2009. The requirement that the defendant seek mental health treatment is a possible condition of probation (supervised release without serving time in prison). While the type of mental health treatment can vary, it typically includes a psychological evaluation, weekly therapy sessions for the duration of the sentence, and potentially a referral to a psychiatrist for medication. Simply comparing outcomes of probationers who are and are not mandated mental health treatment could result in a biased estimate if judges make their sentencing decisions using additional information that is unobserved to the researcher. In particular, while the data provide information on demographics and criminal history, they do not provide information about mental illness, which is likely correlated with both the judge's sentence and recidivism risk. To address this, I use the randomly-assigned judge's propensity to mandate mental health treatment among other cases as an instrument for actually being assigned mental health treatment as a term of probation. This research design has been used in various applications in which a judge, case worker, or other type of administrator has discretion over a randomly assigned caseload.[4] In order for this instrument to provide exogenous variation, the judges must be assigned to cases in a method that is unrelated to factors impacting sentencing and future criminal behavior. In North Carolina, judges are assigned in a rotating pattern which is independent of the individuals in their jurisdiction (Sloan et al., 2016; Silveira, 2017). To verify random assign-ment, I conduct balance tests which show that demographic and criminal background characteristics, while predictive of receiving therapy, are not predictive of the assigned judge's residualized propensity measure.

---

[1] Among them, approximately 22% had a serious mental illness, 37% had a moderate mental illness, and 40% had a mild mental illness (Bagalman and Cornell, 2018).

[2] Any Mental Illness (AMI) is defined by SAMHSA (2022) as having "any mental, behavioral, or emotional disorder in the past year of sufficient duration to meet criteria from the Diagnostic and Statistical Manual of Mental Disorders, 4th edition (DSM-IV), excluding developmental disorders and SUDs [Substance Use Disorders]"

[3] See, for example, Fazel et al. (2010), Grady and Gough (2014), Hiday et al. (2002), and Wüsthoff, Waal, and Gråwe (2014) for treatment as a protective factor in risk of violence; Ziemek (2017) for a review of evidence on treatment in correctional facilities; Green (2020) for a review of the link between mental illness and criminal behavior.

[4] See, for example: Kling (2006), Dobbie, Goldin, and Yang (2018), Leslie and Pope (2017) for other applications in the criminal justice system; Kalish (2023) for the foster care system; Belloni et al. (2012) for eminent domain cases; Autor and Houseman (2010) for evaluation of a welfare program.

The analysis consists of two main parts, first evaluating the impact of mandated mental health treatment on recidivism and then evaluating the cost-effectiveness of mandating mental health treatment for judge-chosen individuals on probation. Using the judge instrument, I find that mandated mental health treatment as a term of probation decreases future crime in both the short and longer term. Three years after conviction, individuals sentenced to mental health treatment are 12.1 percentage points less likely to recidivate. That effect corresponds to a 36 percent decrease in three-year recidivism relative to the mean, and is robust to various specifications of the instrument and treatment. The IV estimate is more than twice as large as the OLS estimate, consistent with judges being more likely to assign treatment to those with underlying mental health issues, who are at a higher risk of recidivism.

In addition, I find that the effect of mental health treatment persists over time, driven by reductions in the first three years. By five years after conviction, offenders are still about 11 percentage points (25 percent) less likely to recidivate. That suggests that mental health treatment is not just shifting when recidivism occurs. Because the average length of probation is about six months, the persistence of the effect means the treatment is effective well beyond the period during which it is mandated. Lacking information on treatment after probation, persistence could be driven by the effectiveness of treatment even after it ends, or instead by the offenders continuing to seek treatment after the mandated period is over. Nevertheless, the persistence stands in contrast with evaluations of cognitive behavioral therapy (CBT) programs in more targeted populations, which find that effects quickly dissipate after the treatment ends (Heller et al., 2017; Blattman, Jamison, and Sheridan, 2017; Arbour, 2021).

To explore heterogeneity in the treatment effect, I estimate two-stage least squares regressions among various subsamples. The results from this exercise suggest that treatment decreases recidivism similarly among different types of crime, first-time and repeat offenders, men and women, offenders of different races and ethnicities, and offenders with different financial resources. The effects range from a 10 percentage point decline (among offenders with a private attorney) to a 20 percentage point decline (among repeat offenders), but are not statistically distinguishable from one another. This analysis shows that offenders with many different characteristics are similarly benefited by mental health treatment. I also test whether treatment has heterogeneous effects on the type of future crime. I find that being mandated to seek mental health treatment is most effective at reducing financial crimes, and is also highly effective at reducing violent and property crimes.

A main identifying assumption underlying this research design is that the sentencing judge only affects recidivism through assignment of mental health treatment. In other contexts the judge has control over more aspects of the sentence, but in North Carolina during this time period, a series of structured sentencing laws mechanically determined the offender's sentence class based on the severity of their offense and the number of prior convictions. Using that information, the analysis sample can be restricted to those who are only eligible for probation. On the other hand, judges do have control over other special conditions of probation. Along with mandating mental health treatment, judges can also require individuals to seek substance use disorder (SUD) treatment. Building on Humphries et al. (2023), Mogstad, Torgovitsky, and Walters (2021), Vytlacil (2002), Mountjoy (2022), and others, I construct a model of judge decision-making that is consistent with the North Carolina sentencing context and with unordered partial monotonicity. Several tests show that the observed patterns in the data are consistent with that monotonicity assumption. The empirical strategy in this paper identifies a margin-specific local average treatment effect, holding the judge's propensity toward SUD treatment constant while varying the judge's propensity toward mental health treatment. This allows me to isolate the effect of mental health treatment from that of SUD treatment.

The instrument variable strategy identifies the local average treatment effect (LATE), which is the causal effect of mental health treatment for those on the margin of being sentenced to treatment as a term of probation. However, as Blandhol et al. (2022) note, the result of two-stage least squares with covariates is often not interpretable as the LATE. It produces an average of covariate-specific LATEs only in the case where the specification is saturated in the covariates; accordingly, I fully interact all covariates as a robustness check of my main results. A related concern is the validity of the monotonicity assumption, which requires that if one judge is ranked as more likely to assign mental health treatment than another judge, that ranking will hold true for all offenders regardless of their individual characteristics. In my main results, I rely on a relaxed version of that assumption by creating my instrument separately for different types of offenses. In addition, as a robustness check I follow Mueller-Smith (2023) and others by creating my instrument separately for many groups defined by a range of demographic and criminal history characteristics and their interactions. The results of both corrections to address covariate bias are qualitatively similar to my main results.

In the second part of my analysis, I turn to quantifying the costs and benefits of mental health treatment among probationers. I carry out the analysis within five broad offense groups, using separate estimates of the effect of treatment and the costs of recidivism for each group. This exercise focuses on the direct effect of mandated treatment on recidivism, but mental health treatment could also improve other aspects of life. Therefore this exercise can be thought of as an underestimate of the total benefit of the treatment. Currently North Carolina expects probationers to cover the costs of mental health treatment themselves; I consider a policy in which the government instead takes on those costs. Such a policy would provide benefits through the reduction of direct and indirect costs of crime at a rate of 5 to 1. I relate the individual's willingness to pay to the net costs to the government via the marginal value of public funds (MVPF) of the policy. Incorporating various aspects of uncertainty from the estimates, I estimate an MVPF of 19.3. This suggests that among probationers, mental health treatment for the marginal individual is highly beneficial for society. Paired with the evidence that treatment effects are fairly homogeneous, the results of this analysis suggest treatment could be similarly cost-effective in a broader context.

The idea that treatment of mental illness could improve more than just symptoms of that illness has been documented by many observational studies in sociology, psychology, and criminology (Hakulinen et al., 2021; Bubonya, Cobb-Clark, and Wooden, 2017; Mojtabai et al., 2015; Burton et al., 2008; Breslau et al., 2008; Patel and Kleinman, 2003). However, causal evidence on the treatment of individuals with poor mental health who have had contact with the United States criminal justice system is mostly limited to evaluations of mental health courts and assisted outpatient treatment. Both programs are focused on individuals who have a prior diagnosis of severe mental illness and are at a crisis point. Though most studies have found reduced criminal behavior after these programs, the group being treated is likely to improve over time regardless, since they are committed at "rock bottom" (Mitton et al., 2007; Frank and McGuire, 2010). In addition, most analysis focuses on short term outcomes measured in the months after contact with mental health courts or assisted outpatient treatment. And few studies provide a discussion of the cost-effectiveness of the programs. I contribute to this literature by presenting evidence on a treatment that is less targeted; offenders are not required to meet thresholds of mental illness severity or even to be diagnosed with a mental illness in order to be sentenced to seek treatment. I evaluate short-term and longer-term effects of mental health treatment, finding that the effects of judge-mandated mental health treatment persist at least three years after the sentence. Importantly, the results of this study also show that bringing about a reduction in crime through mandated mental health treatment is a cost-effective strategy.

My findings are consistent with recent studies of other behavioral interventions in the criminal justice system, but I expand on those studies by presenting evidence on a treatment that explicitly focuses on addressing mental health. A 2002 review by Pearson et al. of behavioral programs administered in prisons identified decreases in recidivism, but concentrated among programs that included a cognitive component rather than just behavioral modifications. That suggests scope for improving mental functioning in order to bring about behavioral change. Most of the interventions they review are carried out in the prison system; in contrast, Heller et al. (2017) evaluate RCTs among low income youth in Chicago and estimate a decrease in recidivism of similar magnitude, between 20 and 30 percent. Importantly, the effect they estimate disappears after the program ends, whereas my estimate persists to at least three years after sentencing. Similarly, Arbour (2021) studies a cognitive behavioral therapy (CBT) program in a Canadian prison that teaches prisoners increased awareness of their thought patterns; he finds that the program decreases recidivism, but the effectiveness dissipates quickly after release for repeat offenders. In contrast, I find that mandated mental health treatment is similarly effective and persistent for both first-time and repeat offenders. I also study adults on probation, an important group that is more than twice the size of the incarcerated population in the United States. Prendergast et al. (2015), Hall, Prendergast, and Warda (2017), Guydish et al. (2011), and Scott and Dennis (2012) show that attempts to incentivize voluntary participation in programs to treat mental health have been largely unsuccessful. My evidence suggests that mandating participation can be an effective strategy to increase use of mental health treatment.

In addition, development economists have used RCTs to study factors related to mental health in many countries within South America, South and East Asia, and Africa. Most of those studies focus on the benefits to economic productivity, finding that antipoverty programs improve mental health which in turn induces economic gains (Ridley et al., 2020). Blattman et al. (2022) specifically identify a decrease in antisocial behaviors like robbery and selling drugs, which persists as far out as ten years. This literature focuses on policies that relax income constraints. Relatedly, Jácome (2020) studies the relationship between losing Medicaid eligibility and crime in the United States, and shows that the detrimental effects are concentrated among individuals with a prior history of mental illness. Consistent with this, Burns and Dague (2023) show that enrollment in Medicaid reduces the likelihood of recidivism and that the effect operates both through financial and health channels. In contrast, I show that even when faced with additional costs, being sentenced to get mental health treatment decreases recidivism. Because in North Carolina the cost of therapy is usually not covered by the court system, mandating mental health treatment introduces a competing force which might make offenders more likely to recidivate. They could violate their probation if they aren't able to attend treatment due to financial constraints. That suggests the estimated effect in this study is likely an underestimate of the effect of mental health treatment, and might be larger in an environment where the responsibility to pay is not on the offender. On the other hand, being required to pay could create deterrence effects that make the treatment more effective for some groups. In this paper I provide suggestive evidence that both are true among different subsets of the probationers studied.

## 2 Context

### 2.1 Probation in the United States

Probation makes up a large part of the criminal justice system. The Bureau of Justice Statistics estimates that 4.2 million adults were on probation in 2009, compared to 1.6 million adults in prison in the United States (West, Sabol, and Greenman, 2010; Glaze and Bonczar, 2010). Probation is a broad term for sentences that do not include incarceration.[5] The probationer is expected to obey all laws and any additional requirements imposed by their judge or probation officer. For example, if probation is the result of a drug crime, submitting to drug testing will likely be a requirement of probation. They are also expected to report regularly to their probation officer, and the frequency and terms of those meetings can depend on both the severity of the offense and the individual judgment of the officer. Breaking the law or not following any additional requirements imposed is considered a violation of the probation, though whether a violation is reported to the court is at the discretion of the probation officer. Once a violation has been reported, a probation violation hearing will be held and the judge will determine whether the terms were indeed violated. Consequences for a violation span from an extension of the probation sentence to the introduction of an active prison sentence, and may also involve paying additional fines.

---

[5]A sentence of incarceration is often referred to as an "active" sentence. In this paper, references to a "non-active" sentence mean a sentence that does not involve incarceration.

Among those on probation, the prevalence of any diagnosable mental illness was 33 percent, twice as high as in the general population. Probationers also reported twice as much unmet need for mental health services – 8.4 percent compared to 4.4 percent in the general population (Feucht and Gfroerer, 2011). Poor mental health can vary widely in intensity, from severe mental illness to occasional symptoms that do not meet the criteria for a full diagnosis. Because Feucht and Gfroerer's estimate of mental illness among probationers does not include individuals who experience poor mental health without meeting the criteria for a mental illness, it may represent an underestimate of the amount of probationers who could benefit from improvements to their mental health. Other studies of mental health interventions in the criminal justice system have focused on high-risk situations, such as creating separate court systems for individuals with serious mental illness; in comparison, a study of probationers likely includes individuals with a larger variation in the degree of poor mental health. By evaluating a more diverse population in terms of their underlying mental health, this paper provides evidence about the effectiveness of treatment beyond the most severe cases.

Mental illness may be related to criminal activity by reducing the ability to make rational, welfare-maximizing decisions. In the context of the criminal justice system, it is correlated with higher risk of committing future crimes and with higher risk of probation violations (Ziemek, 2017; Louden and Skeem, 2011). Mandating mental health treatment for likely-in-need individuals on probation has the potential to mitigate the higher risk of recidivism by treating the underlying mental illness, teaching better decision-making, and supporting longer term behavioral change. In addition, it is difficult to get people to voluntarily participate in effective programs to treat mental health. Prendergast et al. (2015), Hall, Prendergast, and Warda (2017), Guydish et al. (2011), and Scott and Dennis (2012) show that attempts to incentivize participation in community treatment (using financial incentives and intensive case management programs) have not worked. Because it is a requirement of probation, mandated mental health treatment does not suffer from those same implementation difficulties.

When mental health treatment is mandated as a term of probation, it can take many forms. The judge might specify the detailed conditions of treatment, but often they are left to the probation officer to decide. The required components of the treatment commonly consist of an evaluation, counseling sessions, and possible referral to a psychiatrist for medication. Probation officers will often help connect probationers to therapists who work with patients in the criminal justice system. Those therapists use a range of treatment modalities, including but not limited to motivational interviewing, stages of change, cognitive behavioral therapy, reality therapy, Gorski's model of relapse prevention, and moral reconation therapy (Melissa Enoch, personal communication, May 26, 2023). Following the initial evaluation with a licensed practitioner, counseling sessions are generally weekly for the duration of probation, which lasts between six months and a year on average. About half are taking depression, anxiety, or bipolar medications (Tremayne Butler, personal communication, June 16, 2023).

Judges mandate mental health treatment for about five percent of probationers in my sample. Importantly, offenders required to seek mental health treatment are also required to pay for that treatment themselves. Without Medicaid coverage, individuals would typically pay $100 to $300 per session (Melissa Enoch, personal communication, May 26, 2023). The financial pressure this applies could cause individuals to be more likely to violate their probation if they are unable to pay for treatment. If so, mandated treatment might be more effective in an environment in which individuals are not required to cover the cost of treatment by themselves.

The effect of mandated mental health treatment is ambiguous because of these contextual details. Countervailing forces, including the expectation that individuals cover the costs of the treatment and the threat of violating probation if they do not comply, could mechanically result in increased recidivism in response to mental health treatment mandates. In addition, some individuals may respond to mandated mental health treatment with more hesitance than if they were to seek out that treatment freely. On the other hand, the financial cost of mandated treatment could offer an additional deterrence against future crime among those who are able to pay, making the treatment more effective.

## 2.2 Mapping of North Carolina Court System to Empirical Strategy

This paper's empirical strategy exploits variation in assigned sentencing judges' propensities to mandate mental health treatment. There are several characteristics of the criminal court system within North Carolina that make it an appropriate setting for this research design. I first briefly describe the sentencing protocol North Carolina follows, which allows me to limit the analysis sample to only those offenders who will receive a non-prison sentence. I then discuss the two court levels and how judges are randomized within them.

During the time period of this analysis, North Carolina followed a protocol known as structured sentencing. Appendix Figures A1 and A2 depict how, under structured sentencing, different combinations of offense severity and prior convictions determine the class of punishment and the minimum and maximum sentence length. Using this protocol, I restrict the analysis sample to only those who, based on their charged offense, would have been assigned to punishments other than an active prison sentence. The restriction is based on the charged offense rather than the convicted offense to avoid selection on an outcome, since judges could potentially manipulate the specific convicted offense in order to achieve a certain sentence level. By making the restriction to non-active sentences, I can identify the effect of mental health treatment during probation compared to being on probation but not receiving mental health treatment. The estimated effect does not include comparisons between offenders sentenced to probation and those sentenced to prison.[6]

Judge randomization in the North Carolina criminal court system has been used for identification in previous work by Sloan et al. (2013), Sloan et al. (2016), and Silveira (2017). Because the randomization differs between the two court levels, I first provide a brief overview of the different courts. The North Carolina court system is broken into District and Superior Courts, which differ in the severity of crimes they oversee. The District Court hears cases involving misdemeanors and infractions, whereas the Superior Court hears cases involving felonies and appeals. District Court trials are before a judge with no jury, whereas Superior Court trials are before a judge and a jury of twelve. Counties are formed into groups called districts in both courts, and all cases in a district are tried in the county seat. As Appendix Figure A3 shows, by 2009, North Carolina was divided into 42 District Court districts (served by 270 judges) and 50 Superior Court districts (served by 95 judges).

At the Superior Court level, the exogenous variation used to create the judge instrument comes from the way judges are rotated between districts. In addition to the grouping of counties into districts, the Superior Court further groups districts into circuit. The map in Appendix Figure A3 shows the eight circuits present in 2009. Every six months, judges within a circuit move from one district to the next based on the rotation pattern of their circuit.

In the District Court, judges are elected directly to the districts and do not rotate between them. Instead, the exogenous variation for the judge instrument comes from the random assignment of judges into the schedule of trials. The chief judge in the district assigns a judge for each slot from among those available. Each chief judge prepares the schedule for three-month periods (January-March, April-June, July-September, October-December) at a time and aims to assign a judge to the same courtroom for the entire week.

Following the judge assignment mechanism, I create the instrument conditional on court-time fixed effects. Within the Superior Court these are circuit by district by year indicators. Within the District Court they are district by year by day of week by shift indicators to address assignment of judges based on schedule availability. Thus in the Superior Court the strategy compares individuals assigned to different judges due to which judge happened to be stationed in their district at the time they were sentenced, while in the District Court the strategy compares individuals assigned to different judges due to the scheduling of cases by the chief judge. I describe the creation of this measure in detail in

---

[6]There is a third type of sentence in the United States, parole, which differs from probation in that it is assigned to offenders after they have served a prison sentence, often in conjunction with an early release as a reward for good behavior. The Structured Sentencing Act also eliminated parole, so the only offenders serving a non-active sentence are those on probation.

Section 4.

Judges interact with the offender for the duration of the trial, which varies between District and Superior Courts. District Court sentencing trials take only a few hours, while Superior Court sentencing trials can last between one and three days because of the added complexity of the jury trial (Kurtz-Blum, 2021). Even in cases with a jury, only judges determine the imposed sentence, subject to the criteria defined by the structured sentencing guidelines.[7] Judges also interact with offenders when their probation is revoked. In this case, offenders are randomly assigned a judge for their revocation court appearance. In the analysis sample, I identify about 13 percent of offenders who fail their probation.[8] To verify the random assignment of judges in probation violation cases, I calculate the number of offenders who are assigned the same judge for the probation revocation as they are for their original court appearance. I then test whether the frequency of same-revocation-judge assignment (8.9 percent) is different from the probability of being randomly assigned any one judge at the court-district level (7.7 percent). The one-sided test that offenders are more likely to be assigned to their original judge upon revocation has a p-value of 0.28, suggesting offenders are not more likely to be assigned their original judge. This supports the argument that revocation judges are indeed randomly assigned.

## 3 Data

The empirical analysis uses data from the North Carolina court system. Criminal records come from the Administrative Office of the Courts and allow for identification of judge assignment of mental health treatment as well as any future interactions with the criminal court system. Those records are matched with judge schedules compiled from archived documents, which provide necessary information to form the instrument. The resulting dataset is crucial for this project as it provides time variation, enough information to characterize a mental health intervention, and a source of exogenous variation. The rest of this section summarizes the most relevant information about the data; more detailed information about collection and cleaning can be found in Appendix C.

The North Carolina Administrative Office of the Courts (AOC)'s Automated Criminal Infractions System (ACIS) provides information about all District and Superior Court criminal cases. For both courts, the data have information at the case, charge, and defendant level. The charge-level data provide the initial charged offense, the outcome of each charge, and the punishment for guilty charges. Using North Carolina's offense grading system, I assign each charged offense a severity and characterize each case based on its most severe charged offense. The case-level data provide initial arrest date, trial dates, sentencing date, type of attorney, points incurred from prior convictions, and judge presiding over the sentencing case. The defendant-level data provide demographic information, including sex, race, ethnicity, date of birth, and address. Using the defendant's name and date of birth, I create links between cases. I can track their prior and future offenses that occur within North Carolina between 1994 and 2009. This includes charges in different counties and jurisdictions from the current case.

The ACIS identifies the judge initials associated with each sentencing trial. I supplement that with additional information about judge tenures, schedules, and districts using archived documents from the NC Judicial Department. Using the schedules and district maps, I assign judges to their circuit (Superior Court) and district (Superior and District Courts). This information is necessary to form the instrument, which relies on variation in judges within districts.

The main outcome of interest, recidivism, measures whether the offender reappears in the records within three years of the conviction. This measure of recidivism is not specific to incarceration or probation – anyone who enters the criminal justice system again within three years will be identified. In additional analysis I restrict the definition of recidivism to only consider new charges unrelated to probation violations. Doing so addresses the potential mechanical effect of mental health treatment on returning to the criminal justice system due to violations related to not attending treatment or not being able to pay. Defining recidivism within three years covers the period of time during which recidivism is most likely while also balancing sample size for power. As Figure 1 shows, recidivism is most likely within the first year, and recidivism in the first three years makes up about 75 percent of all recidivism that will occur within ten years of the current conviction. In further analysis I investigate how the impact of mental health treatment changes as the definition of recidivism incorporates more years post-conviction.

---
[7]The exception is cases in which the death penalty is an option, at which point the jury is included in the sentencing decision. Because this study focuses on only cases for which the charged offense is mild enough to limit offenders to a non-prison sentence, it is very likely that judges are always the ones determining the sentence.

[8]This number is likely a lower bound due to data quality concerns with that variable.

Crucial to the analysis in this paper, the ACIS data include a field for special conditions of the case which contains additional details about the punishment assigned to guilty charges. I use this field to characterize aspects of the probation sentence, including whether the defendant was required to seek mental health treatment or to seek substance use disorder (SUD) treatment. Although substance abuse is often co-occurring with poor mental health, the two diagnoses are viewed separately by much of the psychology and criminal justice literature. For example, drug courts and mental health courts are two separate potential interventions that target different severely-at-risk individuals in the criminal justice. Those categorized as receiving mental health treatment in my data might also receive SUD treatment, but must receive some kind of treatment that is specific to mental health. In additional analysis I consider alternative definitions of treatment which use a more restrictive definition of mental health treatment or broaden the definition to include more types of treatment.

I observe whether defendants are mandated mental health treatment, but not whether they comply with that mandate. I also do not observe whether defendants seek mental health treatment outside of the mandate. These factors both potentially bias the estimated effect of treatment toward zero by either adding non-treated individuals to the treatment group or by adding treated individuals to the control group. I can use the special conditions field to identify cases that later violate probation and the judges associated with that violation, which identifies some of the individuals who do not comply with the mandate. In Section 5 I show suggestive evidence that indeed, offenders who are assigned mental health treatment are slightly more likely to violate probation, and that my estimate of the effect of mental health treatment is larger when that is adjusted for.

## 3.1 Sample Characteristics

I make two main restrictions to the data when forming the analysis sample. The first requires all defendants to have key information, crucially a judge and information about that judge's scheduled district and circuit. The second restriction limits the sample to those individuals whose only option for punishment was probation. Following the Structured Sentencing laws, offenders are assigned points based on the number of prior sentences and the severity of their charged offense. Those laws then classify defendants into punishment classes based on their points. Certain defendants are only eligible for an active sentence, meaning time in prison; others are only eligible for a community or intermediate sentence, meaning time on probation. I keep only those individuals who were not eligible for an active sentence, representing about 90 percent of misdemeanors and 60 percent of felonies. I also limit cases to those seen during the years the Structured Sentencing laws were in place, from October 1994 to December 2009. Restricting to probation cases in this way, the analysis sample represents about 70 percent of all cases seen during this time period.

The restriction to non-active sentences potentially changes the makeup of individuals in the sample. Appendix Table A1 details these differences, providing summary statistics for the sample before and after implementing the restriction. The sample of probationers looks similar in most ways to the overall sample of criminal court offenders. The most notable difference is that probationers have committed less severe crimes, particularly fewer violent crimes, and have shorter sentences.

Table 1 provides descriptive statistics for the main estimation sample, restricted to observations with demographic and criminal history controls. The sample consists of about 720,000 case-level observations. I divide the sample into two groups based on whether they were mandated to seek mental health treatment. The table shows that those who were mandated to seek mental health treatment (Column 2) were about one third less likely to recidivate in the three years following their sentence. Though this provides suggestive evidence that mandated mental health treatment could reduce future crime, there are clear imbalances in the comparison groups that could lead to bias in a naive comparison of means. The group assigned mental health treatment differs in their demographic characteristics: they are less likely to be Black and slightly more likely to be female. They differ in their criminal history: they are more likely to be first-time offenders and have fewer prior arrests. They also are more likely to have private attorneys and to be assigned SUD treatment.

The most common type of crime among probationers is traffic and public order offenses, as Figure 2 shows. Figure 3 shows that mental health treatment is also most frequently assigned among traffic and public order offenses, followed closely by drug and alcohol offenses.[9] It is least likely among financial and fraud offenses - only about two percent are assigned treatment there, compared to almost eight percent among traffic and public order offenses. Treatment is also more common among sex offenders and offenses related to domestic violence. These differences illustrate the potential need for a source of exogenous variation to form a comparison that is not selected on characteristics, like severity of offense, that might be correlated with the likelihood of recidivism.

## 4 Method

This paper estimates the effect of judge-mandated mental health treatment on future criminal activity. A naive method to estimate that effect would use the baseline regression given by Equation 1, where $Y_{ict}$ is the outcome of interest for individual $i$ in case $c$ in year $t$.

$$Y_{ict} = \tilde{\beta}_0 + \tilde{\beta}_1 \text{MHT}_{ict} + \epsilon_{ict} \tag{1}$$

Judges decide whether to mandate mental health treatment based on information they have about defendants at the time of the trial. If the information they use is correlated with future criminal activity, then the simple regression based on Equation 1 would produce a biased estimate of the effect of treatment. Controlling for a rich set of demographic characteristics and the criminal history of defendants would recover the causal effect of treatment on recidivism if those characteristics made up the full set of information the judge used to make their decision. In practice, I will show that indeed, the coefficient $\tilde{\beta}_1$ changes in magnitude and the adjusted $R^2$ increases considerably when a vector of controls $\mathbf{X}_{ict}$ is included in the regression. However, I address concerns that even after controlling for those characteristics, judges might still have more information than the data provides.[10] To identify the causal effect of mandated mental health treatment on future recidivism, I instrument for $\text{MHT}_{ict}$ with $\texttt{JudgeLikelihood}_{ict}$, the judge's propensity to assign treatment among other offenders.

I estimate a specification following Equation 2, replacing the actual mental health treatment sentence with the predicted sentence $\widehat{\text{MHT}}_{ict}$ from Equation 3. Assuming judges are monotonically more lenient across all characteristics of convicted individuals, this strategy identifies the local average treatment effect (LATE), which is the causal effect of mental health treatment for those on the margin of being sentenced to treatment as a term of probation.

$$Y_{ict} = \beta_0 + \beta_1 \widehat{\text{MHT}}_{ict} + \beta_2 \mathbf{X}_{ict} + \gamma_{ct} + \varepsilon_{ict} \tag{2}$$
$$\text{MHT}_{ict} = \alpha_0 + \alpha_1 \texttt{JudgeLikelihood}_{ict} + \alpha_2 \mathbf{X}_{ict} + \tilde{\gamma}_{ct} + \epsilon_{ict} \tag{3}$$

In Equations 2 and 3, exogenous variation comes from judge assignment within districts, as discussed in detail in Section 2. Accordingly, all analysis controls for court-time fixed effects ($\gamma_{ct}$). Within the Superior Court those are circuit-district-year fixed effects; within the District Court they are district-year-shift-day-of-week fixed effects. Various specifications control for the same demographic and criminal history characteristics of the defendant ($\mathbf{X}_{ict}$) as in the OLS specification described earlier.

---

[9] Figures 2 and 3 only report the main offense. For example, if drug and alcohol abuse is more commonly a secondary offense, could be contributing to higher rates of mental health treatment among other offenses like public order violations.

[10] In particular, the data do not have information about the individual's mental health, which could both influence the judge to assign mental health treatment and make recidivism more likely.

Those characteristics include race, ethnicity, sex, a polynomial in age, region of the United States, type of attorney, and indicators for: first-time offender, a prior offense in the last year, offense type (violent and property, financial and fraud, traffic and public order, drugs and alcohol, and miscellaneous), and a sex offender. Certain specifications also include county-level controls for the per capita number of psychologists, emergency visits due to mental health, Medicaid and CHIP recipients, residents without insurance, residents with bachelor's degree or more, residents living in poverty, and violent crimes committed. Because counties are subsets of court districts, those controls allow for exploration of differential access to care among individuals with the same set of possible judges as a potential mediating factor in the effect of treatment.

## 4.1 Multiple Aspects of Sentence

Judges have discretion over more than the assignment of mental health treatment. This introduces concerns about the exogeneity of the instrument in the empirical design. If the instrument, which is measuring the judge's underlying tendency toward mental health treatment, is correlated with the judge's tendency toward assigning other aspects of the sentence, then movements in the instrument will affect recidivism through more channels than solely mandated mental health treatment.

As described in Section 2, situating this analysis during the period of time when North Carolina used Structured Sentencing rules out many areas over which judges might usually have discretion. The decision to incarcerate, for example, is governed by the severity of the convicted offense and the number of previous convictions. To assuage concerns that judges might still indirectly affect whether offenders are incarcerated by manipulating the convicted offense class, I test whether judges who differ on mental health treatment assignment differentially manipulate the convicted offense. This test regresses either an indicator for whether the charged and convicted offense are the same or an indicator for whether the charged and convicted offense result in the same sentence class on the judge's propensity to assign mental health treatment. The estimates suggest that having a judge who is more or less inclined toward mental health treatment is not correlated with manipulation of the sentence class thresholds. Nevertheless, analysis is instead based on the charged offense, which is determined earlier in the judicial process by the arresting officer and at times the prosecutor. The conviction dimension is explored further in Section 5.6.

The Structured Sentencing laws also govern the length and intensity of probation. For example, the least severe offense class (Class 3 Misdemeanor with no prior points) restricts sentences to "community" (non-supervised) probation for one to ten days. The most severe offense class among those that do not have the possibility of incarceration (Class I Felony with 5-8 prior points) restricts sentences to "intermediate" (supervision by a designated probation officer) probation for four to eight months. Conditional on sentence class, judges do have minimal discretion over the exact length of probation within the requirements. The magnitude of that discretion changes depending on the sentence class: for the least severe sentence, the most judges could do is cause the offender to be under unsupervised probation for nine more days. Other sentence classes are characterized by wider ranges of allowed length. Following a similar strategy as in the case of testing potential manipulation of the convicted offense classes, I find that judge tendencies toward mental health treatment are not correlated with the actual sentence length assigned. In addition, some sentence classes allow for either community or intermediate (unsupervised or supervised) probation. Testing the relationship between judge tendencies toward mental health treatment and sentence class results in an estimated correlation of -.03 (standard error .052). These results support that judge tendencies toward mental health treatment are not correlated with whether probation is supervised or unsupervised.

On the other hand, judges have open discretion over assigning other types of treatment programs. In particular, judges can require a convicted individual to seek treatment for drug abuse and addiction. A damaging relationship with drugs is very common among individuals in the criminal justice system. Using 2007-2009 National Inmate Surveys, Bronson and Berzofsky (2017) estimate that 58 percent of state prisoners in the United States and 63 percent of sentenced jail inmates met the criteria for drug abuse. In comparison, the rate of drug abuse in the general population at the time was five percent. The previous literature evaluating programs that address drug dependency in the criminal justice system is mixed. It provides suggestive evidence of improvements in recidivism rates as a result of the introduction of drug courts, which pull those with drug dependency issues into a separate court process that focuses on rehabilitation. Very little exploration has been done on the cost efficiency of such courts, which can be quite expensive to implement and run (Doleac, 2023).

In the context of this paper, drug programs present an empirical challenge in that judges have discretion over both assigning mental health treatment and assigning substance use disorder (SUD) treatment. To address the challenge I build on the literature on the identification and estimation of treatment effects in the presence of multiple treatment alternatives (e.g. Carneiro, Heckman, and Vytlacil (2011), Mogstad, Torgovitsky, and Walters (2021), Heckman and Pinto (2018), Lee and Salanié (2018), Mountjoy (2022)). My approach most closely follows work by Humphries et al. (2023) on identification of multiple treatments in the judge instrument setting. Carneiro, Heckman, and Vytlacil (2011) first pointed out that in the case of multiple instruments, not controlling for the other instruments incorporates their effect on the outcome into the estimator if they covary with the focal instrument. Mogstad, Torgovitsky, and Walters (2021) show that using one instrument while conditioning on the others preserves the Imbens and Angrist (1994) interpretation of LATE as a weighted average over compliers. They consider a scenario with multiple instruments for one treatment and introduce partial monotonicity, which requires that monotonicity be satisfied for each instrument holding the others constant. Humphries et al. (2023) expand that result to multiple endogenous treatments, showing that depending on the model of judge decision-making, certain margin-specific treatment effects can be identified. In their setting they impose what Mountjoy (2022) calls unordered partial monotonicity, which combines Mogstad, Torgovitsky, and Walters's partial monotonicity with a version of Heckman and Pinto's (2018) unordered monotonicity. Unordered monotonicity requires that shifting an instrument moves all agents uniformly toward or against each possible choice.

Adapting the notation of Humphries et al. to the setting of this paper, define $T_t$ as a binary indicator of assignment to treatment $t \in \{M, D\}$ where $M$ refers to mental health treatment and $D$ refers to substance use disorder treatment. Consider the continuous judge propensity instruments $z_t$ for each treatment. Then unordered partial monotonicity with respect to mental health treatment can be defined as:

$\forall z_M, z'_M, z_D \in supp(Z)$ and fixing $z_D$, either (1) or (2) is true:

1. $T_M(z'_M, z_D) \geq T_M(z_M, z_D)$ and $T_D(z'_M, z_D) \leq T_D(z_M, z_D)$

2. $T_M(z'_M, z_D) \leq T_M(z_M, z_D)$ and $T_D(z'_M, z_D) \geq T_D(z_M, z_D)$

Because unordered partial monotonicity is margin specific, $UPM(z_M|z_D)$ refers to fixing the judge's propensity toward SUD treatment and varying the propensity toward mental health treatment. The $UPM(z_M|z_D)$ condition requires that moving from a judge who is less likely to assign mental health treatment to a judge who is more likely, holding constant the judge's SUD treatment propensity, weakly induces some individuals into mental health treatment. In addition, the second part of the assumption requires that increasing the judge's mental health treatment propensity does not move individuals out of mental health treatment. Taken together, these result in a restriction that induces complier flows into only one treatment. Notice that $UPM(z_M|z_D)$ is not the same as $UPM(z_D|z_M)$; it is possible that only some margins will be identifiable. The implication of this assumption, as Mogstad, Torgovitsky, and Walters (2021) and Carneiro, Heckman, and Vytlacil (2011) also stress, is that $z_D$ must be conditioned on in Equations 2 and 3 to identify the effect of mental health treatment on recidivism from variation in $z_M$.

Several models of judge decision making could be consistent with unordered partial monotonicity, and many models would not be. For example, if the judge were to make each decision completely independently without relying on any of the same information for either choice and the necessary information for each choice were unrelated to that for the other, then unordered partial monotonicity would trivially hold because $T_M$ would only depend on $z_M$, $T_D$ only on $z_D$, and $z_M$ and $z_D$ would be uncorrelated. In this setting, that is unlikely to be an accurate model of how the judge decides on sentencing. However, here I describe a model of judge decision making, based on Vytlacil (2002) and informed by characteristics of the North Carolina judicial system and the data, which is consistent with the $UPM(z_M|z_D)$. For some functions $\pi_t$ for treatment $t \in \{M, D\}$, where $\pi_t$ are measurable and nontrivial functions of $z_t$,

$$T_D = \mathbb{1}\{\pi_D(z_D) \geq U_D\} \tag{4}$$
$$T_M = \mathbb{1}\{\pi_M(z_M, z_D) \geq U_M\} \tag{5}$$

In this latent index selection model, consider two sources of unobserved-to-the-researcher factors, $U_D$ and $U_M$. The offender's relationship with drugs is summarized by $U_D$ and the offender's mental health status is summarized by $U_M$. When a judge determines an offender's sentence, she first evaluates whether that offender has a drug dependence issue. If not, she has no reason to assign SUD treatment; if so, she has some likelihood of assigning treatment based on the intensity of the evidence of drug dependence. After making this determination, she moves on to consider other potential terms of probation. Information on drug abuse is likely relatively easy for the judge to learn based on the facts of the case and interaction with the offender in the courtroom, but information about the offender's mental health status might be harder to accurately determine. Because of that, the judge makes her decision about recommending mental health treatment based both on her perception of the offender's mental health and drug abuse. This model allows for a situation in which the judge views an offender with a history of drug abuse, decides that offender is more likely to have a mental illness, and orders them to seek mental health treatment. Holding $z_D$ constant and varying $z_M$ moves individuals across only one threshold, inducing compliers into or out of mental health treatment. Thus this framework is consistent with $UPM(z_M|z_D)$. However, holding $z_M$ constant and varying $z_D$ could induce movement along multiple dimensions, since both $T_D$ and $T_M$ depend on $z_D$. Thus this framework does not support $UPM(z_D|z_M)$, and the causal effect of SUD treatment, holding mental health treatment constant, cannot be identified from the model.

Unordered partial monotonicity results in testable implications. In particular, the characteristics of individuals who are assigned SUD treatment, once controlling for the judge's SUD treatment propensity, should not vary by the judge's mental health treatment propensity. This can be seen in Equation 4, where being assigned SUD treatment only depends on the judge's drug propensity, not on the judge's mental health propensity. Following Humphries et al. (2023), I test this implication by first predicting recidivism using a vector of individual characteristics and then regressing that predicted recidivism on the judge's propensity toward mental health care, controlling for the judge's propensity toward SUD treatment, in the subset of the sample who received SUD treatment. The p-value for the test of whether mental health treatment is predictive of the characteristics of those assigned SUD treatment is 0.84. The results of this test suggest that the mental health treatment propensity is not related to the individual characteristics that predict recidivism among SUD treatment recipients. The test supports the $UPM(z_M, z_D)$ assumption holding in the context of this study.

To summarize, the approach here relies on the independence of $T_D$ from $z_M$. Then conditioning on $z_D$ when estimating the effect of $T_M$ instrumented with $z_M$ identifies the local average treatment effect of mental health treatment, following Mogstad, Torgovitsky, and Walters (2021), Carneiro, Heckman, and Vytlacil (2011), Humphries et al. (2023), Mueller-Smith (2023), and others. Notice that $T_D$ is in the error term and is correlated with $T_M$ because both depend on $z_D$. Controlling for $z_D$ breaks that correlation. This exercise can be thought of in the framework of Haavelmo causality: the causal parameter is defined using hypotheticals that fix all other inputs and vary one to determine outcomes (Haavelmo, 1943, 1944; Heckman and Pinto, 2015; Pearl, 2015). Because judge assignment of mental health treatment and SUD treatment are endogenous, there is no model of a natural experiment that varies mental health treatment assignment without also varying SUD treatment assignment. However, holding the judge's SUD propensity constant allows for conceptualizing the estimated effect of mandated mental health treatment as the partial derivative of a structural model.

## 4.2 Instrument Creation

This paper follows a growing literature that constructs an instrument based on exogenously assigned adjudicators' decisions on other cases during the same period. In this section I first summarize the general method the literature has taken to creating this instrument, then describe the changes this study makes to incorporate recent theoretical advances in the literature.

The instrument is a measure of the tendency for a randomly assigned judge to assign individuals to some outcome or group. In other papers that has been pretrial detention, incarceration, foster care homes, and more.[11] In the context of this paper, the judge assigns individuals to seek mental health treatment as a term of their probation sentence. The previous judge instrument literature commonly uses the phrase "judge leniency," but in this context it is less clear what is more lenient and what is more strict. Assigning mental health treatment could be seen as more strict, giving offenders more requirements they have to satisfy in order to fulfill probation. It could also be seen as more lenient, giving offenders a chance to seek help in changing their behavior rather than trying to work through it alone. I instead refer to the instrument as the "judge propensity" or "judge tendency" toward mental health treatment.

The measure is constructed by first residualizing the relevant judge decision on the vector of court-time fixed effects that govern the randomization of the judges. As described in Section 2, the court-time fixed effects ensure comparison of offenders who were at risk of being assigned to the same possible judges. In the Superior Court, they include the circuit (to which the judges are elected), the district (where they serve, and are rotated between), and the year. In the District Court, they include the district (to which the judges are elected and where they serve), year, day of week, and shift. The resulting residual is averaged over all cases seen by that judge, excluding the individual's case. Finally, the average among all judges is subtracted from the average for that judge, so that the resulting instrument variable measures how much a given judge deviates from the average judge's sentencing behavior. In the context of this paper, positive values of the instrument characterize judges who are more likely to assign mental health treatment. The leave-out measure of judge propensity is constructed by regressing the judge's sentence ($\texttt{MHT}_{icj,t}$) on the vector of court-time fixed effects ($\gamma_{c,t}$) in Equation 6 and then taking the mean of the residual, subtracting any cases related to that individual, in Equation 7. Here $n_{j,t}$ refers to the number of cases the judge oversees in a year and $n_{ij,t}$ refers to the number of cases the judge oversees for individual $i$.

$$\widehat{\texttt{MHT}}_{icj,t} = \texttt{MHT}_{icj,t} - \gamma_{c,t} \qquad (6)$$

$$\texttt{JudgeLikelihood}_{ij,t} = \frac{1}{n_{j,t} - n_{ij,t}} \left( \sum_{k=0}^{n_{j,t}} \widehat{\texttt{MHT}}_{kj,t} - \sum_{l=0}^{n_{ij,t}} \widehat{\texttt{MHT}}_{lij,t} \right) \qquad (7)$$

This describes the basic form of the judge instrument. Section 5.6 discusses alternative forms of the instrument that address various potential concerns and incorporate recent literature exploring the exogeneity and homogeneity properties of the judge instrument. For the main specification, following Leslie and Pope (2017) and others, I calculate the residualized mean separately by felony and offense group.[12] Calculating the instrument within types of crime relaxes the monotonicity assumption, which requires that if one judge is ranked as more likely to assign mental health treatment than another judge, that ranking will hold true for all offenders regardless of their individual characteristics. The relaxed assumption instead requires that the judge ranking be preserved within, for example, all violent felonies, and within all public order misdemeanors, but not necessarily between those two groups. Section 5.6 discusses further relaxations of this assumption.

---

[11]See, for example: Kling (2006), Dobbie, Goldin, and Yang (2018), Leslie and Pope (2017), Kalish (2023), Belloni et al. (2012), Autor and Houseman (2010)

[12]The offense groups are the same as those referenced in Figure 2: violent and property, financial and fraud, traffic and public order, drugs and alcohol, and miscellaneous.

Figure 4 depicts a histogram of the main residualized measure. Positive values represent those judges who are more likely to assign therapy. The distribution is slightly skewed right, with a minimum value of -0.33, a maximum value of 0.98, and a standard deviation of 0.03. This implies that moving from a judge three standard deviations below the mean propensity to a judge three standard deviations above the mean propensity increases the probability of mental health treatment by about 18 percentage points. Most judges assign mental health treatment fairly rarely, but very few never assign it. Figure 5 plots the percent of cases judges assign to treatment, organizing the judges by how often they assign mental health treatment. The 50th percentile judge assigns mental health treatment less than ten percent of the time. On the right tail of the judge distribution, there are judges who assign mental health treatment more than sixty percent of the time.

### 4.3 Balance

For this strategy to identify a causal effect, the classic instrument variable assumptions must hold. The judge tendency to sentence offenders to seek mental health treatment must be correlated with actual receipt of a mental health treatment mandate and not related to other factors that predict recidivism. In addition to the histogram of the instrument, Figure 4 also plots the residualized rate of therapy assignment against the judge propensity measure. It represents a graphical depiction of the first stage: as the propensity measure increases, judges are more likely to assign therapy. This provides evidence that the judge instrument is predictive of mental health treatment, and indeed the F-statistic from the first stage is quite large, over nine hundred.

The randomness of judge assignment is crucial to the validity of the estimation strategy. If judges are exogenously assigned, the instrument should not be related to other factors that predict recidivism, satisfying the exclusion restriction. In order to verify random assignment, I carry out three tests of whether judges with higher propensity are assigned offenders with different characteristics.

The first test provides evidence that the judge is a stronger predictor of the offender's mental health treatment sentence than of the observable characteristics of offenders with mental health treatment. I first form the predicted probability that an offender would be assigned mental health treatment on the basis of their demographic and criminal history characteristics and the court-time fixed effects. The predicted probability is then regressed on a full set of judge indicators, controlling for the court-time fixed effects. I compare the F-statistic from that regression to the F-statistic from a regression of the actual mental health treatment assignment on the set of judge indicators. The F-statistic for the predicted probability is 0.49 (p-value 0.48), implying that there are no differences in the ex-ante characteristics of offenders who appear before different judges. The F-statistic for the actual treatment is larger and significantly different from zero (p-value 0.01), which provides evidence that judges do matter for determining the offender's assignment to mental health treatment.

For the second test, I regress the judge propensity measure and mental health treatment assignment on demographic characteristics and criminal history, conditional on court-time effects, and test the joint significance of these controls. Table 2 presents the results of this test. The F-statistic for the mental health treatment mandate is 405, with p-value approximately zero. The large F-statistic shows that the individual characteristics are highly predictive of mental health treatment assignment. In contrast, the F-statistic for the judge propensity instrument is 2. The small size indicates that the offender's characteristics are not significant factors in predicting the judge's propensity. Though the associated p-value is also small (0.001), the size of the estimated coefficients on the characteristics that predict judge propensity are not economically significant. The two characteristics of concern are having a private attorney and being a sex offender. Those characteristics increase the judge propensity toward mental health treatment by one one-hundredth and three one-hundredths of a standard deviation, respectively. In addition, many of the characteristics that are predictive of being mandated mental health treatment have no impact on the judge propensity measure. Black probationers are less likely to be sentenced to mental health treatment, while first-time offenders and female probationers are more likely. None of those characteristics are predictive of the judge's propensity to assign treatment. This test suggests that the measure of judge propensity is generally balanced across demographic and criminal background controls, but identifies the potential for some sorting across judges by sex offenders and individuals with private attorneys.

For the final test, I compare the first stage with and without including a standardized version of the judge instrument. Table 3 presents these results. The judge instrument is predictive of mandated treatment: having a judge with a one standard deviation higher propensity toward mental health treatment increases the likelihood of being assigned mental health treatment by about 2.4 percentage points. In contrast, the judge's tendency toward SUD treatment has a much smaller relationship with mandated treatment, increasing the likelihood by about 0.1 percentage points. Comparing the final two columns of the table shows that including the instrument has little effect on the predictive power of the controls, but does increase the $R^2$. Similarly, comparing the second and third columns shows that adding controls has little effect on the magnitude of the estimated relationship between the instrument and treatment, but does increase the $R^2$. Those comparisons suggest that the connection between the instrument and mandated mental health treatment is not related to the individual demographic and criminal history characteristics of the probationers.

Taken together, these three tests suggest that the process of case assignment to judges is conditionally random. However, because of the potential for sorting related to sex offenders and offenders with private attorneys, I control for the vector of observed demographic and criminal history characteristics described above in the main analysis specification. Section 5.2 provides evidence that the effect of treatment is similar among offenders of different demographic characteristics, including those with private versus public attorneys. That suggests any sorting on defendant characteristics is not driving the main estimate of the impact of mental health treatment on criminal activity.

## 5 Results

### 5.1 The Effect of Mental Health Treatment on the Incidence of Future Crime

Table 4 presents the main results for the sample that combines District and Superior courts.[13] I estimate Equation 2 instrumenting for mental health treatment sentence with the leave-out residualized measure of judge propensity to assign treatment. The outcome of interest is recidivism, measured as an indicator of ever returning to the criminal court system within three years of the current offense. I limit my main analysis to cases tried between 1994-2006 so that each offender has the chance to reappear in the data by three years since the trial. All analysis, unless otherwise stated, controls for court-time fixed effects and the judge's tendency toward SUD treatment. All reported standard errors are clustered at the court-time level, and F-statistics are robust (Kleibergen and Paap, 2006).

The first column of Table 4 presents a simple form of the baseline OLS specification.[14] Omitted variables could bias the estimated effect of mandated mental health treatment in either direction. On the one hand, judges might only send individuals to therapy if they seem to have a good chance of using it to their advantage, biasing the estimate away from zero (positive bias). On the other hand, judges might only send individuals to therapy if they look to be struggling more, biasing the estimate toward zero (negative bias). In practice, controlling for demographic and criminal history characteristics in column two has a large impact on the magnitude of the estimated effect, dropping it in half from ten to five percentage points. This is consistent with judges choosing to assign mental health treatment based on observable characteristics that are associated with lower likelihood of recidivism, such as being a first time offender.

---

[13]In combined-court specifications, all explanatory variables are interacted with an indicator for court.

[14]The estimate from the first column (0.10) differs from the difference in means in Table 1 (0.12) because it controls for court-time fixed effects.

If the vector of included controls were to span the full support of the information judges use to assign mental health treatment, then the estimate in column two should identify the causal effect of mental health treatment on recidivism. However, instrumenting for mandated mental health treatment with the associated judge propensity results in estimates that are larger in magnitude than estimates from the OLS with controls specification, suggesting that the included controls did not encompass the judge's full information set. The larger magnitude of the IV estimates means that the unobserved factors were biasing results toward zero in OLS. This is consistent with judges choosing to assign mental health treatment to individuals who are more likely to have a mental illness, which is related to a higher risk of recidivism. Columns three through six present the estimate from the IV specification with various included control variables. Column six presents the main specification, which controls for the judge's propensity toward SUD treatment as well as the demographic, criminal history, and county characteristics described in Section 4. I find that being assigned to seek mental health treatment decreases the likelihood of being rearrested within three years by about 12.1 percentage points. For reference, 34 percent of the convicted probationers over the sample period returned to the court system within three years of being sentenced. Thus the effect corresponds to a 36 percent decrease in three-year recidivism relative to the mean.

Overall, these results suggest that there is a group of offenders for whom mental health treatment is an effective means of deterrence from future crime. The results are consistent with recent evidence on the interaction between behavioral programs, financial access to healthcare, and recidivism. Jácome (2020) finds that low-income young men who lose Medicaid eligibility are 22 percent more likely to be involved in the criminal justice system, and among men with mental illness that number rises to 40 percent. Arbour (2021) finds even larger effects in his study of a Canadian prison's cognitive behavioral therapy (CBT) program: a 58 percent decrease in recidivism compared to prisoners who do not participate in the program. Heller et al. (2017) identify a 20 to 30 percent decrease in crime due to a CBT intervention among low-income Chicago youth. The effect I estimate falls in line with those estimates, though the referenced studies are estimating shorter-term effects. They focus on recidivism during and in the months following the program, whereas my definition of recidivism includes years after the average probation sentence has ended. Heller et al. (2017), for example, find that the effect they estimate disappears after the program ends, whereas my estimate persists beyond the mandatory period of treatment.

The results from this study are also consistent with studies of alternative courts for individuals with mental health or substance abuse concerns. A review of 27 drug court evaluations by the GAO (2005) found rearrest reductions of about 10 to 30 percentage points. An evaluation of the Calgary Diversion Program, a mental health court, by Mitton et al. (2007) found that it was correlated with decreased court reappearances of between 25 to 48 percent. Both programs focus on different contexts than that of this paper. Individuals routed through drug courts do not necessarily have a mental illness, whereas the Calgary Diversion Program required individuals to have a serious mental illness. In contrast, the mental health of the population studied by this paper likely falls somewhere in between those of the drug court and mental health court. In addition, this paper specifically studies the treatment of mental health, rather than the different court processing of those with mental illness. Nevertheless, the cited studies provide an anchor to which to compare this study's estimates, and show that they are in line with previous work on mental and behavioral interventions in the criminal justice system.

The main estimate should be thought of as an intent-to-treat estimate in the sense that there is no information on how many offenders assigned to mental health treatment follow through and take up treatment. For a rough proxy of whether the offender follows through with treatment, I estimate the effect of mental health treatment on whether the offender fails probation in Table 5. Probation can be violated for many other reasons, such as failing to pay court fines, traveling somewhere forbidden, or committing another crime. And reported violations are in part determined by probation officers, who differ in their strictness. With those caveats, I find suggestive evidence that offenders sentenced to mental health treatment are more likely to fail their probation. Indeed, when I estimate the effect of mental health treatment on three-year recidivism where I instead treat probation violation cases as *not* committing a future crime, my estimate is slightly larger. I discuss probation violations further in Section 5.6.

## 5.2 Heterogeneity in the Effect of Mental Health Treatment on Recidivism

I explore heterogeneity in the treatment effect by estimating two-stage least squares regressions among various subsamples defined by demographic characteristics, criminal history, and characteristics of the court district. In general, my results are suggestive of little treatment effect heterogeneity. Offenders with many different characteristics benefit similarly from mandated mental health treatment.

Figure 6 plots the estimated effect of mandated mental health treatment on three-year recidivism within each group of individual characteristics. I find negative effects of treatment on recidivism for most subgroups, though some estimates are imprecise. The point estimates range from a decrease of about 10 percentage points to a decrease of about 20 percentage points. The estimates provide suggestive evidence of slightly larger effects among repeat offenders, women, younger offenders, and offenders with a public attorney, but the estimates are not statistically different from each other. The estimate of the effect of mental health treatment among Hispanic offenders is particularly imprecise, with a confidence interval that spans from a decrease of 45 percentage points to an increase of 15 percentage points. The imprecision is most likely due to the small number of observations - only about two percent of the sample is Hispanic, consistent with the demographics of North Carolina in the early 1990s.[15] Overall, the results in this figure suggest that mental health treatment is similarly effective among many types of individuals.

The results regarding the heterogeneous effects of mental health treatment among types of offenders differ from previous literature in one particularly notable way. Some prior work (see, for example, Arbour (2021)) finds that first time offenders are more responsive to interventions that decrease recidivism. In contrast, in the context of mandated mental health treatment, the effect is not concentrated among first time offenders but instead proves similarly, or if anything more, beneficial for repeat offenders.

The homogeneity of the treatment effect provides some assurance that the judges are not being assigned based on characteristics of defendants. For example, one potential concern could be that first-time offenders, who are less likely to commit future crimes, are assigned certain judges who are more prone toward mandating mental health treatment. In that case, the effect of mental health treatment among first-time offenders should be larger; instead it is indistinguishable from the effect among repeat offenders. Similarly, defendants with a private attorney might be less likely to commit future crimes because they have additional support and more financial security. If their attorneys were able to bargain for a judge who is less prone to mandating mental health treatment, the effect of treatment would be biased toward zero. Instead, the estimated effect of mental health treatment is similar among defendants with private and public attorneys. Heterogeneity of the effect would not necessarily indicate a violation of the exclusion restriction, but the lack of heterogeneity at least rules out some violations.

I also explore whether different kinds of offenders are more or less impacted by the potential for mandated mental health treatment to result in a probation failure due to lack of attendance or inability to pay. Figure 7 shows that there is more evidence for heterogeneous treatment effects when the outcome is failing probation. The strongest evidence is in the comparison of private and public attorney. The positive relationship between mental health treatment and probation violations is driven entirely by individuals with a public attorney, who are almost 20 percentage points (confidence interval [10, 30]) more likely to fail probation as a result of being assigned mental health treatment. By contrast, being assigned mental health treatment has no effect on probation violations among individuals with a private attorney. This evidence is consistent with financial constraints being an important mediator in the implementation of mental health treatment. I explore further evidence of the importance of

---

[15]While the population of Hispanics in North Carolina experienced enormous growth starting in the 1990s, only 1.2 percent of the population was Hispanic in the 1990 Census. That number had increased to 4.7 percent by 2000. (Census, 1990, 2001)

financial constraints in Section 5.3.

I find the most evidence of heterogeneous treatment effects among different offense types. Table 6 presents the results from splitting the sample by whether the individual was charged with a misdemeanor or a felony. The estimates suggest mandated mental health treatment is more effective among individuals who were arrested for more serious crimes. The estimated effect of mental health treatment for those individuals is about 18 percentage points (50% of the mean among felonies), whereas the estimated effect among individuals who committed misdemeanors is only about 10 percentage points (30% of the mean among misdemeanors). Table 7 shows the results from estimating the main equation separately by the class of the convicted offense. These estimates suggest that mental health treatment is most effective among offenders who have been arrested for violent, property, or financial crimes. Treatment is also effective among offenders who have been arrested for drug and alcohol crimes. The particularly strong effect of mental health treatment among probationers charged with financial crimes could be explained by financial deterrence effects. Being required to pay for treatment may be seen as a particularly salient punishment to those who are convicted of financial crimes. In contrast, mental health treatment is much less effective among those who commit traffic and public order violations. I estimate an increase of 2.5 percentage points in recidivism among that offense group, but the estimate is not distinguishable from zero. Overall, these results show that individuals charged with more severe offenses experience more of a benefit from mental health treatment, especially when the crimes are felonies.

## 5.3 The Effect of Mental Health Treatment on Recidivism over Time

The main results focus on recidivism within three years. To understand how the effect of mental health treatment evolves over time, I explore several specifications. The results suggest that the effect of mental health treatment persists in the first three years after trial, during which offenders are at the highest risk of recidivism, and then dissipates. As Figure 1 shows, recidivism is most likely within the first year, and recidivism in the first three years makes up about 75 percent of all recidivism that will occur within ten years of the current conviction.

To understand how the effect of mental health treatment differs over time, I consider two measures of recidivism. One is cumulative: it characterizes recidivism up to $x$ years after the trial. When $x = 3$, this corresponds to my main measure of recidivism. Panel A of Figure 8 shows the results from the cumulative measure looking up to five years after trial. Mental health treatment is similarly effective as time elapses, with point estimates between a six and twelve percentage point decrease in recidivism. The confidence intervals are large enough that the different estimates are not statistically distinguishable from one another. Appendix Table A2 presents the point estimates corresponding to the figure. Appendix Figure A5 presents a version of this exercise based on estimates from an OLS regression of recidivism on mandated mental health treatment. Although the magnitudes of the point estimates differ from the IV estimates (as discussed in Section 5.1), the OLS results show similar trends in the effect over time.

The relatively flat trend over time could be consistent with several different explanations, as it incorporates changes in the effect, in the frequency of the dependent variable, and in the sample.[16] In Appendix Figure A4, I repeat the same exercise using a balanced sample of cases from 2004 and earlier. The results are less precise, but the pattern is similar to that in the main sample, with estimates hovering around 10 percentage points. The comparison with the balanced panel suggests that the results are not driven by changes in the cases included in the sample. Turning to the frequency of the dependent variable, Panel B of Figure 8 shows the results as a percent of the mean in the relevant time period. In part because the cumulative rate of recidivism increases over time, the estimated effect in percent terms decreases over time from about 30 percent to about 25 percent of the mean.

---

[16]The sample changes because the measure of recidivism requires the trial to have been at least $x$ years before the last year in the data.

To understand the mechanisms underlying the decrease in the effect in percent terms over time, I turn to a second measure of recidivism over time. This measure divides time after the trial into single-year periods, so that $x = 3$ would correspond to only recidivism between two and three years after the trial. The results from this exercise are shown in Figure 9. Panel A suggests that mental health treatment has the largest estimated effect on recidivism in the first year since trial, when the offender is significantly more likely to commit a new crime. The estimate in the first year since trial suggests mental health treatment reduces recidivism by about 9 percentage points. The results in the following two years are smaller and less precise, suggestive of a decrease in recidivism by about 3 percentage points for each successive year. However, as Panel B shows, the effect of mental health treatment measured as a percent of mean recidivism is quite similar across the first three years. The evidence from this exercise suggests that the effects of mandated mental health treatment persist for three years after the trial and dissipate after. Returning to the percent effects in Panel B of Figure 8, comparison to the second measure of recidivism suggests the persistence of the effect over five years is driven by reductions in recidivism in the first three years.

Because probation spans between five to nine months on average, the persistence of the effect during the first three years after trial suggests that treatment is effective even beyond the period during which it is mandated. It also suggests that mandated mental health treatment is not just shifting back when recidivism occurs, as would be the case if the estimated effect changed signs after the initial years. Lacking information on take up of treatment, I cannot say whether the persistence is driven by skills learned and a reduction in mental illness which continues after treatment ends, or instead by the continued use of treatment after the mandated period is over. Nevertheless, the persistence stands in contrast with evaluations of other behavioral programs which find that effects mostly disappear after the end of the program. Blattman, Jamison, and Sheridan (2017) show that CBT effectiveness depreciates over time, and the cognitive-behavioral intervention among low income youth studied by Heller et al. (2017) lost effectiveness quickly after the youth exited the program. The interventions studied in previous research differ from the intervention studied in this paper in several ways, as discussed in Section 1. The focus on treating underlying mental illness, the freedom of individuals to choose a specific type of therapy, and the setting among individuals on probation could all contribute to the longer persistence of the estimated effect of treatment.

To understand potential mechanisms driving the effectiveness of mental health treatment over time, I compare the trajectory of the effect for individuals with private versus public attorneys. Viewing private attorney as a rough proxy for the offender having a higher socioeconomic status, these results are consistent with the effectiveness of treatment being related to access to resources. Figure 10 presents the percentage point and percent estimates over time for individuals with a private attorney and individuals with a public or waived attorney. The effect of mental health treatment in percentage points increases slightly over time for individuals with a private attorney, whereas for individuals with public attorneys the effect starts to diminish after three years since trial. When incorporating differences in the likelihood of recidivism over time for these two groups of offenders, the difference in the trends is amplified: the effect in percent terms diminishes more sharply over time for those with a public attorney, whereas it increases after the first year for those with a private attorney and then remains flat for the following four years. Offenders with private attorneys could be more able to follow through with treatment during probation and to continue treatment after the mandated period because of the monetary cost associated with treatment. Indeed, these results suggest that treatment remains effective for years after probation among individuals with a private attorney, whereas it diminishes in effectiveness over time for individuals with a public attorney. Although the point-in-time estimates are, for the most part, not statistically different across the subsamples, the different shape of the trends over time is consistent with the importance of financial access to mental health treatment as a moderator for the effectiveness of the treatment. This evidence is also consistent with a potential financial deterrence effect of the monetary cost of treatment. Individuals who can pay may nonetheless have a distaste for doing so, and may be incentivized to avoid future contact with the criminal justice system in order to escape future financial penalties.

## 5.4 The Effect of Mental Health Treatment on the Type of Future Crime

The main results focus on how mental health treatment affects the extensive margin of criminal behavior, measured by the appearance of a future criminal conviction. In this section, I consider various ways of characterizing the intensive margin of crimes committed in the future. I first exchange the dependent variable for a measure of how many crimes were committed within the three years following the trial. Table 8 presents these results. Including cases that do not recidivate, the average number of future crimes is about 0.5.[17] I estimate that being mandated mental health treatment results in about a 43 percent decrease in the number of crimes committed.

In addition to the number of future crimes, I evaluate the length and intensity of future punishment and the severity of future crimes. Table 9 presents the unconditional and conditional results for the effect of mandated mental health treatment on the average future sentence length and the likelihood of a future active sentence (incarceration). Consistent with the negative effect of mandated treatment on the extensive margin of recidivism, I show that it also reduces the overall length of sentences and likelihood of an active sentence, by 55 days (57%) and 6.2 percentage points (43%) respectively. Probationers assigned mental health treatment also appear to serve shorter sentences and fewer prison sentences conditional on having committed a crime, though conditioning on recidivism reduces the sample such that the estimates are imprecise. Turning to the characteristics of the offense, I estimate the effect of mental health treatment on different classes of crime in Table 10. These results are unconditional, and so incorporate both the overall effect of mandated mental health treatment on recidivism as well as any differential effects on separate types of crimes. I find that being assigned mental health treatment is most effective at reducing future violent or property and financial or fraud crimes (an 11 and 18 percentage point reduction, respectively). The estimates suggest that mental health treatment has a smaller effect on future traffic offenses, which are likely less impacted by underlying poor mental health. Finally, I estimate the effect of mental health treatment on future misdemeanors and future felonies in Table 11. I find that mental health treatment is more effective at reducing future felonies, with no statistically distinguishable effect on reducing misdemeanor crimes.

These results are consistent with evidence from previous studies of behavioral interventions that have identified effectiveness at reducing violent crimes (see, for example, Blattman et al. (2022)). Even more than violent crimes, mental health treatment is most effective at reducing financial crimes. Jácome's (2020) study of Medicaid provision among low-income young men came to a similar conclusion, showing that losing Medicaid eligibility lead to increases especially in financially-motivated crimes. Overall, my analysis of the effect of mental health treatment on the intensive margin of crime suggests that treatment is effective by reducing the number and severity of crimes in addition to the incidence of any crime.

## 5.5 Returning to the Contribution of Substance Use Disorder (SUD) Treatment

In the above analysis, I control for the judge's propensity toward SUD treatment in order to identify the margin-specific local average treatment effect of mandated mental health treatment. As discussed in Section 4.1, this approach follows Humphries et al. (2023), who build on Mogstad, Torgovitsky, and Walters (2021) and Carneiro, Heckman, and Vytlacil (2011), among others. Here, I investigate how important controlling for the drug propensity is in practice. Table 12 presents results analogous to Table 4, but without controlling for the judge's propensity toward SUD treatment. The specification including demographic, criminal history, and county controls estimates a 14.1 percentage point decline in recidivism as a result of being assigned mental health treatment. That estimate is about two percentage points larger than the main estimate controlling for drug propensity, though the difference is not statistically significant. The direction of the bias suggests that ignoring the dependence of mental health treatment assignment on the judge's propensity toward SUD treatment overestimates the effectiveness of mental health treatment. Estimates from the literature evaluating drug courts suggest that SUD treatment reduces certain types of future crime (Mitton et al., 2007; Frank and McGuire, 2010). Since mental health treatment and SUD treatment positively covary, the 14.1 percentage point estimate is likely picking up some bias from the benefits of SUD treatment in reducing recidivism. However, the bias is not large, consistent with the relatively small correlation between mental health treatment and SUD treatment.

---

[17]Conditional on committing at least one crime in the future, the average number is 1.8.

## 5.6 Robustness in the Estimated Effects

I carry out several robustness exercises to ensure that the results are not sensitive to the choice of specification or sample. In this section I discuss those exercises, grouped into four categories: the design of the exogenous variation, the monotonicity assumption of the instrument, judge discretion over conviction, and the choices made when creating the analysis sample and forming the outcome and treatment variables. Overall I find that the results are not driven by these choices. Across all robustness exercises, the results show that mandated mental health treatment contributes to a lower likelihood of criminal activity.

### 5.6.1 Setting of Exogenous Variation

The design of the exogenous variation in this study relies on judges being assigned as-good-as-randomly at the court-time level. The randomization level varies between the District and Superior court systems. In the main specification I define the court-time fixed effects governing the randomization level as district-year-shift-day-of-week groups in the District court, and circuit-district-year groups in the Superior court. I consider two alternate definitions of the court-time level.

The baseline specification addresses the potential that certain judges are, for example, more available in the mornings, when one type of case happens to be frequently scheduled. The first alternative specification removes time variables, using only the district or circuit-district to form the groups. That would be consistent with no scheduling constraints that are correlated with certain types of cases or offenders. In the Superior court, judges are randomly assigned at the district level but based on a pool of judges determined by the circuit to which the district belongs. Accordingly, the baseline specification controls for the circuit as well as the district. The second alternative specification removes circuit, grouping both District and Superior courts by the district alone. Appendix Table A3 presents the results of this exercise. Changing the court-time level, which affects both the variables on which the instrument is residualized and the level of clustering for the standard errors, has a minimal impact on the estimated effect of mental health treatment. Compared to a baseline estimate of 12.1 percentage points, the first alternative level results in an estimate of 13.9 percentage points, and the second alternative results in an estimate of 15.2 percentage points.

The construction of the instrument also depends on the amount of judge information incorporated. Following the previous literature on the judge instrument strategy, the baseline construction of the instrument uses information from all cases the judge sees over the sample period, removing information from cases related to each probationer. Here I discuss two groups of alternative formulations to address concerns about exogeneity and to incorporate recent methodological work on judge instrument design.

First, I address concerns that the judge's underlying propensity toward mental health treatment might be changing over time and that change might be correlated with the offenders the judge sees. I consider alternate formulations of the instrument that use only cases from the judge's first year, that omit cases from future years, that only use cases in the current year, and that only use cases in three-year groups. Appendix Table A4 presents results from these specifications. Using only the judge's first year, the instrument is much weaker and the estimate is not distinguishable from zero. However, other approaches to creating the instrument result in strong first-stage relationships and similar estimates of the effect of mental health treatment on recidivism.

Second, following Frandsen, Leslie, and McIntyre (2023), I create a cluster-jackknife version of the instrument, in which I leave out not only the individual's case but all other cases in the individual's randomization cluster (for example, all cases in District 3 in 1997 on Monday mornings in the District Court). This addresses concerns that the types of individuals grouped into that cluster might differ in meaningful ways from the types of individuals in another cluster, and that those differences would be incorporated into the instrument, injecting endogeneity. Using this instrument, I estimate a 12.7 percentage point decrease in recidivism. The slightly smaller F-statistic compared to my baseline specification, and the point estimate that is slightly more different from the OLS estimate do not rule out that the baseline instrument might have been biased by the inclusion of variation in the types of individuals in a cluster, but the similarity of the estimates suggests that it was not a large source of bias.

### 5.6.2 Monotonicity of the Instrument

The validity of the empirical design for identifying a local average treatment effect (LATE) relies on the assumption of monotonicity (or homogeneity) of the instrument. Section 4.1 characterized a type of monotonicity that allows for the identification of margin-specific treatment effects, holding constant the judge's propensity toward SUD treatment. In this section, I discuss how the monotonicity assumption changes when incorporating control variables.

The instrument variable strategy identifies the margin-specific local average treatment effect, which is the causal effect of mental health treatment for those on the margin of being sentenced to treatment as a term of probation, holding constant the judge's propensity toward treatment. However, when the specification includes covariates, the LATE interpretation does not necessarily apply. As Blandhol et al. (2022) note, the linear IV instead estimates treatment effects for both compliers and always-takers, where some always-taker effects are negatively weighted. Two-stage least squares with covariates produces an average of covariate-specific LATEs only in the case where the specification is saturated in the covariates. Furthermore, in the presence of heterogeneous nonlinear treatment effects, even nonparametrically controlling for covariates is not sufficient to identify a causal effect; the first stage must also be monotonicity-correct. The previous literature using the judge instrument design has partially addressed this second point by creating the instrument within groups of offenders. In the main specification, those groups are formed by the felony status and class of the offense. Here, I fully address the monotonicity-correctness by expanding the definition of the groups within which the instrument is formed. In addition, I address Blandhol et al.'s first point by fully saturating the model in the covariates.

Appendix Table A5 presents results estimated with a "rich covariate specification" in which all covariates are interacted. The specification includes single, double, and triple combinations of race, ethnicity, sex, first time offender, whether there was a previous charge in the last year, sex offender, six offense groups, felony, age, court, year, season, day of week, and shift. The table also presents estimates from a specification in which the instrument is interacted with the covariates in the first stage in order to reproduce the direction of monotonicity conditional on the covariates. Interacting the instrument with covariates produces the same result as creating it within many groups of individual and case characteristics, Mueller-Smith's (2023) approach to relaxing monotonicity. The new monotonicity assumption then requires that if one judge is ranked as more likely to assign mental health treatment than another judge, that ranking will hold true for all offenders of a certain type defined by those groups of characteristics. For example, among all Black first-time offenders, one judge must uniformly assign treatment more frequently than another judge.

In practice, interactions between the covariates and instruments result in a large number of potential instruments and many potential controls. In order to avoid overfitting or many-instruments bias, I use Lasso with k-fold cross-validation to partial out those instruments with little explanatory power for predicting mental health treatment and those controls with little explanatory power for predicting recidivism. Following Chernozhukov et al. (2018), I implement a doubly-robust procedure that guards against model selection mistakes and requires only one model to be correctly specified. It relies on Neyman orthogonality combined with sample splitting so that the estimators of the nuisance parameters do not contribute to the limiting distribution of the parameter of interest. Of the 73 possible instruments, Lasso chooses 13. It exclusively chooses interaction terms, particularly interactions of race with other demographic characteristics and of offense groups with criminal history characteristics. The results in Table A5 are estimated using the Lasso-selected instruments and control variables. Correctly saturating the model in covariates and instruments has little effect on the magnitude of the estimated effect of treatment on recidivism. While the specifications with saturated instruments have smaller F-statistics, the results are qualitatively similar to estimates using only the within-offense-and-felony-groups instrument, suggesting that incorporating differences by crime types is sufficient to address any monotonicity violation.

### 5.6.3 Judge Discretion in Conviction

Judges have discretion over whether to convict a defendant, which could introduce bias to the estimates by threatening the exogeneity of the instrument. The design of the empirical strategy requires that the judge's tendency toward mental health treatment must only impact recidivism through its relationship with the assignment of an offender to mental health treatment. If the judge's tendency toward mental health treatment is correlated with their tendency toward convicting, then ignoring that conviction tendency would introduce endogeneity into the instrument. In practice, the correlation between those two tendencies of the judge is very small, at around 0.07. Nevertheless, I explore two alternatives to the main analysis specification. Appendix Table A6 presents those results.

The first group of two columns presents results from estimating Equation 2 on a sample restricted to only those probationers who were convicted. This approach eliminates concerns that the comparison group includes both individuals who were sentenced to probation without mental health treatment and those who were not convicted. However, the sample is selected on an outcome of the judge's decision set, which introduces concerns about selection bias. Results presented in the second group of two columns instead rely on the main analysis sample, but treat the judge conviction tendency like the SUD treatment tendency, conditioning on the judge's likelihood of convicting other offenders in the sample. In both alternative specifications, the resulting estimate of the effect of mental health treatment on recidivism is slightly larger than (but not distinguishable from) the baseline estimate, at about 14 percentage points. The conclusion from this analysis is that any bias introduced by the judge's tendency toward conviction is small and does not change the implications of the results.

### 5.6.4 Analysis Sample and Treatment Variable

Many choices are made when building the analysis sample from the raw court files. Here I discuss several potentially important decisions, and show that the results are robust to alternative choices.

I first address the variable identifying whether an offender is required to seek mental health treatment. The variable is formed by searching for key words in a long string of special conditions and notes that are attached to each case. Appendix Table A7 addresses concerns that the results could be sensitive to which key words are included in the definition of mental health treatment. In order to isolate the differences from changing the mental health treatment variable definition only, this table presents estimates from the specification without controlling for the judge's SUD treatment propensity. The baseline estimate from that specification is a 14.1 percentage point decrease in recidivism due to mandated mental health treatment. Table A7 considers various ways to restrict the key words that define mental health treatment, and shows that the estimated effect remains similar across those definitions. When the definition is more restrictive, the estimated effect is about the same, at 13.4 percentage points. When the definition is less restrictive, the estimated effect is slightly higher, at 15.2 percentage points. When some SUD treatment key words are excluded from the definition, the estimated effect is also larger, at 16.5 percentage points. All estimates have strong first stages, and are well within one standard deviation of the baseline estimate. The preferred specification takes a moderate restrictiveness toward what is considered mental health treatment, which corresponds to a fairly conservative estimate of the effect of treatment.

I now turn to exploring the definition of the SUD treatment variable. Appendix Table A8 presents estimates from the main specification which controls for the judge's propensity toward SUD treatment, and which uses the baseline definition of mental health treatment. The table considers definitions of SUD treatment that vary the restrictiveness of the key words that are included. The baseline estimated effect here is a reduction in recidivism by 12.1 percentage points. Using two less restrictive definitions, the estimate becomes 9.6 and 9.3; using a more restrictive definition, the estimate is slightly higher, at 12.7. These estimates all have strong first stages and none of the effect sizes are statistically distinguishable from one another. The preferred specification again takes a moderate restrictiveness toward what is considered SUD treatment.

The main results explore the effect of mental health treatment conditional on the judge's propensity toward SUD treatment. Another way to conceptualize treatment is as the combination of both special conditions under the judge's control. Appendix Table A9 presents estimates of the effect of being assigned mental health treatment or SUD treatment. This specification implies a different model of the judge's decision-making. Here the judge jointly chooses to assign some sort of treatment (could be counseling, drug rehab, or both) or to assign no treatment, and that decision is dependent on one underlying factor that is unobserved by the researcher. Instrumenting with the judge's propensity to assign any treatment isolates variation that moves some offenders into receiving a treatment. I estimate that being assigned any treatment decreases recidivism by 12.9 percentage points, very similar to the baseline estimate of the effect of mental health treatment. This exercise provides suggestive evidence that the estimate of the effect of mental health treatment is not sensitive to the specific model I impose on judge behavior.

An alternative view more in the style of Heckman and Pinto (2018) would be to characterize treatment as four mutually-exclusive and exhaustive categories. Abbreviating mental health treatment as MHT and substance use disorder treatment as SUDT, the categories are: (1) MHT alone, (2) SUDT alone, (3) MHT and SUDT, and (4) neither MHT nor SUDT. Instruments for these four treatments can be created using the judge's tendency toward each specific category. For example, the instrument for being assigned MHT alone would be the judge's tendency to assign MHT alone among all other offenders in the sample. Appendix Table A10 presents results from this approach, where "neither" is the left-out category. Estimating each of the treatment categories separately suggests that MHT alone is similarly effective as MHT paired with SUDT, with estimated 16.2 and 15.8 percentage point declines in recidivism, respectively. SUDT alone is slightly less effective, associated with a decline of 11.1 percentage points. However, the results including all three treatments together show that the effect is concentrated in MHT alone. The specification with the three treatments suggests that MHT alone, compared to no treatment (the left-out category), reduces recidivism by 15.1 percentage points. The estimates for MHT and SUDT, and for SUDT alone, are not statistically distinguishable from zero. The comparisons from this exercise suggest that the effect of mental health treatment on future crime is operating primarily through the channel of mental health treatment rather than through an interaction with SUD treatment.

The last major choice related to sample definition is the characterization of the outcome variable. For the majority of the analysis, recidivism is defined by whether the offender appears in the data with a later trial date. That definition includes probation violations that lead to a future court appearance. However, it is not clear whether violating probation should count as recidivism. Here, recidivism is instead defined to exclude cases in which the future trial is a probation revocation case. Table 5 presents those results. It first shows suggestive evidence that being assigned mental health treatment slightly increases the likelihood of having a probation violation. Because seeking mental health treatment adds an additional requirement to the probation terms, it could mechanically cause more violations. If probationers are unable to pay for counseling, or do not have the time or access to transportation, they could have their probation revoked. It is unclear how often this happens in practice, as probation officers have discretion to give warnings rather than an official charge of a violation. I find that mental health treatment increases the likelihood of violating probation by 2.6 percentage points, but that estimate is not statistically significant. The next two columns of the table show that the estimated effect of mental health treatment on recidivism is slightly larger when the measure of recidivism does not include probation violations. I estimate a 12.9 percentage point decrease in non-violation recidivism. This exploration of the characterization of recidivism suggests that the preferred specification results in a conservative estimate of the effect of treatment. In a scenario in which the onus of payment were not on the probationer, mental health treatment could potentially be more effective because it might not induce mechanical violations related to financial barriers. However, it is possible that being responsible for payment makes individuals more inclined to participate in the treatment. In evaluating the effect of mental health treatment, the preferred specification counts violations in recidivism, since it characterizes the total effect of treatment under the current mandated environment.

# 6 Policy Implications

The preceding analysis suggests that mandating mental health treatment for a heterogeneous group of individuals on probation would decrease their criminal involvement over a fairly long horizon. In this section I contextualize those results by considering whether the benefits of treatment outweigh the costs of providing it. In North Carolina, probationers are generally required to pay for their own mental health treatment even if it is court-mandated. For the following cost-effectiveness exercises, I first carry out an accounting of the costs and benefits associated with mandated treatment. I then discuss assumptions under which the estimates from this paper can be applied to consider a policy that would instead pay for mental health treatment for the judge-chosen individuals.

I compare the cost of providing treatment with the benefits, which include lower judicial costs, lower social costs to victims, and lower costs to offenders from fewer future criminal justice system interaction. Importantly, I consider the benefits associated with reductions in crime separately by type of crime, since the studied population of probationers is more likely to commit less severe crimes. Ignoring the characteristics of the crimes would result in an overestimate of the benefits of paid provision of mental health treatment. Appendix B describes the approach taken here in more detail and provides ranges for each estimate discussed below. All monetary amounts referenced are in 2022 dollars.

## 6.1 Costs of Providing Mental Health Treatment

To calculate the per-person costs of mental health treatment, I rely on Medicaid reimbursement tables. The main cost of the policy comes from the provision of counseling and psychiatric medication. I define treatment as one initial intake evaluation combined with weekly counseling sessions for the duration of probation. I include the average cost to Medicaid of daily antidepressants scaled by the likelihood of using those medications.[18] On average, providing treatment for one individual would cost the government $3,042. I make two adjustments to that initial estimate. First, I include the marginal excess burden due to deadweight loss if all those funds were raised through an increase in taxes. Second, I remove the costs associated with the approximately 36 percent of probationers who would have already been eligible for and would have taken up Medicaid. That fraction reflects the amount of probationers who both meet the low income threshold and are parents or disabled, since North Carolina did not expand Medicaid eligibility beyond those groups. Combining the two adjustments, my final estimate of the costs of treatment for one individual is $3,233 in 2022 dollars.

## 6.2 Benefits of Providing Mental Health Treatment

To calculate the per-person benefits from the reduction in future criminal propensity, I use separate estimates of the effect of mental health treatment on different classes of crimes. I estimate Equation 8 for future offenses of violent and property, financial and fraud, traffic and public order, drugs and alcohol, and miscellaneous types. The estimation procedure follows the main specification, but here the outcome is an indicator for the individual's future offense being of a certain type and the comparison is to not committing a future crime.[19] These five crime categories are the same groups within which the instrument is created, as described in Section 4.2.

---

[18]Psychiatric medication is used by about half of the individuals seen by counselors who work in the criminal justice system, based on conversations with several of those counselors.

$$\texttt{Offense}_{ict+1} = \beta_0 + \beta_1 \widehat{\texttt{MHT}}_{ict} + \beta_2 \mathbf{X}_{ict} + \gamma_{ct} + \varepsilon_{ict} \tag{8}$$

Assigning a monetary value to the costs of the criminal justice system is a difficult task. On the one hand, information can be gathered on the financial outlays toward operating the courts, police, jails and prisons, probation, and other aspects of the judicial system. On the other hand, the cost of crime to victims and society as well as the detrimental effects of incarceration on offenders are much less straightforward to value. I pull from an important literature characterizing the costs of crime in order to form my cost estimates. I consider the judicial costs, sourced from Hunt, Anderson, and Saunders (2017) and McCollister, French, and Fang (2010); the loss of tax dollars due to decreased productivity of offenders while incarcerated, sourced from McCollister, French, and Fang (2010); the social costs, sourced from Heaton (2010), Hunt, Anderson, and Saunders (2017), and McCollister, French, and Fang (2010); and the costs to the offenders, sourced from McLaughlin et al. (2016). Social costs include both direct losses to victims as well as indirect losses due to feeling unsafe, etc. Whenever possible, I use cost information disaggregated by crime type and the associated estimate of the effect of mental health treatment on future crimes of that type. I then aggregate the cost estimates weighting by the relative prevalence of each offense group among repeat offenders. On average, the costs avoided by providing mental health treatment total $16,131.

### 6.3 Combining Costs and Benefits

I first consider a simple accounting exercise comparing the costs to the offender versus the total benefits through reductions in crime. I make an adjustment to the costs to the offender calculated above, since that estimate gives the cost to the government of providing treatment through Medicaid. Mental health treatment is much more costly out-of-pocket – about twice as costly as listed in Medicaid reimbursement tables. When thinking about offenders' willingness to pay for treatment themselves, the relevant figure would be $5,411.29. Still, comparing the cost to the offender with the benefits ($16,131) shows that mandated mental health treatment is much more financially beneficial than it is costly.

However, many of those benefits accrue to society overall rather than to the offenders themselves. Therefore, I consider a policy that would shift the burden of payment to the government. In order to rely on the same estimated benefits from the reductions in crime, I make two simplifying assumptions about behavioral responses. I assume that both the effect of mandated mental health treatment and the assignment of offenders to treatment by judges do not depend on who pays. Because the causal effect estimated in this paper is that of *being mandated to seek* mental health treatment, the financial burden is part of the treatment. I have shown suggestive evidence in Section 5 that for more financially disadvantaged individuals, being required to pay induces more probation violations, which suggests the effect of mandated mental health treatment would be larger if the financial burden were shifted off of those probationers. On the other hand, I have also shown evidence consistent with a financial deterrence effect on more financially advantaged individuals, which suggests the effect could be smaller if the financial burden were taken away. This policy exercise could be viewed as imposing that any effects of the financial aspect of treatment average out to zero in the population overall. Alternatively, a policymaker might prefer to view this exercise as applying only to those more financially disadvantaged probationers, in which case any behavioral response to the change in payment incidence would render these an underestimate of the cost-effectiveness of the policy.

Figure 11 presents a graphical depiction of the magnitudes of the costs of treatment to the government and benefits of treatment through the reduction in crime. Comparing the costs relative to the benefits, I find that paying for mental health treatment would about equalize with the benefits earned from the reduction in direct costs of prison alone. Incorporating societal costs of crime results in a benefit-to-cost ratio of about 5 to 1. The high ratio suggests that among probationers, mental health treatment for the marginal individual is highly beneficial for society.

---

[19]In other words, I estimate recidivism in the sample of offenders who either commit a future crime of that type or do not commit a future crime.

Another way to evaluate the cost-effectiveness of the policy is through the marginal value of public funds (MVPF), which measures the amount of revenue that can be delivered to the relevant beneficiaries per dollar of government spending on the policy (Hendren and Sprung-Keyser, 2020). The MVPF is the ratio of the individual's willingness to pay to the net costs of the policy to the government. In the case of mental health treatment among probationers, my estimate of the net costs to the government ranges from -$790 to $2,721. The large range comes from two sources, uncertainty due to the estimated effect of treatment and uncertainty due to the disagreement in the literature on the magnitude of the costs of crime. If the savings due to reductions in crime are greater than the cost of the program, then the net cost to the government is negative and the marginal value of public funds is infinite (as long as the willingness to pay is positive). In the case of the lower bound cost estimate, the MVPF of mental health treatment among probationers is infinite. To form the MVPF associated with the upper bound estimated net costs to the government, I turn to estimating the individual's willingness to pay.

To form an estimate of the individual's willingness to pay, I consider both offenders and a representative member of society. The offender benefits by avoiding the negative effects of incarceration, including lost earnings, increased risk of sexual abuse and mortality, and broken family connections. Whether their case is dismissed, they are sent to prison, or they are put on probation, all offenders benefit from the reduced likelihood of shouldering court fees and fines – just the conviction-related court fines alone average $869 for misdemeanors and $1,628 for felonies (Rafael, 2023). The representative member of society benefits in many ways related to the decreased likelihood of being a victim of a crime or a family member of an incarcerated person. Tangible benefits include reduced likelihood of the loss of property and medical treatment. Intangible benefits, much harder to measure, include reductions in fear of crime and the psychological effects of being a victim of crime. I quantify these benefits using the estimates from Section 6.2 for the total social costs and costs to offenders weighted by the effect of treatment on future crime. The willingness to pay for reductions in crime associated from mental health treatment is estimated at $14,526 (with a range between $4,069 and $31,650). Finally, combining the willingness to pay with the net cost to the government, I estimate a MVPF of 19.3, with a range between 1.5 and infinity. That the lower bound estimate of the MVPF is still greater than one provides strong evidence of the cost-effectiveness of the policy.

The findings from the traditional cost-benefit analysis and the marginal value of public funds exercise confirm that spending one dollar on mental health treatment for probationers returns more than one dollar to the beneficiaries of the policy. For comparison, Jácome (2020) estimates an MVPF of 1.8 for a policy related to reducing crime through the provision of health insurance. Her policy involved expanding Medicaid eligibility for two years to all low-income young men in South Carolina. The policy I consider here has much lower costs than Jácome's, which likely contributes to the larger estimated MVPF. Drug courts are another related policy with much higher costs, at about $21,000 per person (Prins et al., 2015). Studies have found potential benefits of these courts, but those benefits are smaller than the estimated effect of mental health treatment on recidivism (see Deschenes, Turner, and Greenwood (1995); Gottfredson, Najaka, and Kearley (2003); Prins et al. (2015)). However, holding the distribution of future crimes in the sample constant, the drug courts would need to be about twice as effective as mental health treatment in order to return an MVPF greater than 1.

## 6.4 Areas for Further Research

I view the cost-benefit analysis carried out here as an underestimate of the cost-effectiveness of mental health treatment among probationers for several reasons. These cost-benefit exercises focus only on the effect of mandated mental health treatment on criminal behavior. If receiving mental health treatment makes it easier for probationers to find and keep employment or achieve more education, then the benefits of mandated therapy would be even larger. In addition, the estimates in this section are based on benefits due to reductions in recidivism in the three years since trial, but the results in Section 5.3 suggest that recidivism risk in later years remains lower for probationers who are sentenced to mental health treatment. The true lifetime benefits of mental health treatment are therefore likely larger than calculated here.

This exercise is also limited in that it takes the assignment of treatment as given. Other potential policies could involve expanding assignment of treatment to more probationers, to other types of offenders, or even to individuals outside the criminal justice system. Because individuals on probation are by definition committing less serious offenses (in order to be assigned to probation by the Structured Sentencing laws), the estimated monetary value of the benefits of reduced crime are much smaller than they could be. For example, the social cost of violent crimes like homicide, rape, and assault is nearly $1 million, but only about five percent of repeat offenses in the analysis sample are violent crimes. In a population in which more serious offenses were committed, the benefits of treatment could be even larger. However, the benefits also depend on how effective mental health treatment is in those settings, which this paper cannot answer. The results in this paper, finding that mental health treatment is similarly effective across many demographic and criminal history characteristics, suggest that treatment could be effective among individuals in different scenarios as well. They also suggest limited scope for targeting the provision of treatment to specific types of offenders, since I do not identify any individual characteristic that is relatively more predictive of the effectiveness of the treatment. However, this paper has shown that probationers are a group for which judge-targeted treatment proves a cost-effective policy to reduce crime.

# 7 Conclusion

This paper evaluates the causal impact of mental health treatment on individuals' likelihood of committing a future crime. Using randomized judge assignment to North Carolina criminal cases, I find that mandated mental health treatment as a term of probation decreases future crime in both the short and longer term. I estimate that being assigned to seek mental health treatment decreases the likelihood of three-year recidivism by about 12 percentage points, or 36 percent. That effect persists over time: by five years after conviction, offenders are about 11 percentage points less likely to recidivate. My estimation strategy addresses the interaction of mental health treatment with another aspect of probation, substance abuse disorder treatment. I show that, with some assumptions about the model of judge decision-making, I can recover a margin-specific treatment effect. I also explore different dimensions of heterogeneity, finding that treatment is similarly effective in many subgroups. I show that the effect of mental health treatment is not concentrated among a certain class of offenses, though the evidence in this paper suggests that mental health treatment is particularly effective at reducing more serious offenses. I validate these results in a series of robustness checks, all which suggest that mandated mental health treatment contributes to lower criminal propensity.

Taken together, the results suggest that mandating mental health treatment as a term of probation can improve the transition of convicted criminals into civilian life. I quantify the costs and benefits of a policy paying for mental health treatment for all judge-chosen probationers, using separate estimates of the costs and the effects of treatment among offense groups. I estimate that the benefits from the reduction in crime would more than recover the costs spent to provide mental health treatment, at a rate of about 5 to 1. The associated marginal value of public funds is 19.3. Among probationers, mental health treatment for the marginal individual is highly cost-effective for society.

Almost half of jail inmates in 2012 had a history of mental health problems (Bronson and Berzofsky, 2017). Individuals with mental illness can struggle to make rational, welfare-maximizing decisions, making traditional incentive-based strategies for crime deterrence less effective. Mental health treatment helps align their preferences with their actions; it can teach strategies to avoid high-risk situations, improve decision-making, support long-term behavioral change, and correct the perceived costs and benefits of behaviors. The findings of this study illustrate the potential of mandated mental health treatment as a beneficial and cost-effective alternative to those traditional crime deterrence strategies.

Table 1: Offender Characteristics by Mental Health Treatment Sentence

|  | All | MHT | No MHT |
|---|---|---|---|
| Age | 30.53 | 32.58 | 30.43 |
|  | (10.15) | (10.66) | (10.12) |
| Female | 0.20 | 0.21 | 0.20 |
|  | (0.40) | (0.41) | (0.40) |
| Black | 0.45 | 0.35 | 0.46 |
|  | (0.50) | (0.48) | (0.50) |
| Hispanic | 0.02 | 0.03 | 0.02 |
|  | (0.15) | (0.16) | (0.15) |
| Sex Offender | 0.01 | 0.03 | 0.01 |
|  | (0.10) | (0.17) | (0.10) |
| First Time Offender | 0.61 | 0.70 | 0.60 |
|  | (0.49) | (0.46) | (0.49) |
| Private Attorney | 0.31 | 0.48 | 0.31 |
|  | (0.46) | (0.50) | (0.46) |
| Assigned MH Treatment | 0.05 | 1.00 | 0.00 |
|  | (0.21) | (0.00) | (0.00) |
| Assigned SUD Treatment | 0.04 | 0.37 | 0.02 |
|  | (0.20) | (0.48) | (0.15) |
| Recidivism within 3 Years | 0.34 | 0.23 | 0.35 |
|  | (0.47) | (0.42) | (0.48) |
| Failed Probation | 0.06 | 0.08 | 0.05 |
|  | (0.23) | (0.27) | (0.23) |
| Observations | 727,184 | 33,886 | 693,298 |

Note: This table presents summary statistics for the main estimation sample. The sample is divided into two categories: those who were assigned mental health treatment and those who were not. Observations are required to have non-missing control and treatment variables and to have been seen at least 3 years before the end of the data so that recidivism can be measured. I designate First-Time Offender based on when offenders first appear within the sample period beginning in 1994, which could misclassify offenders who were arrested in years prior to 1994. "SUD" stands for Substance Use Disorder. Source: 1994-2009 North Carolina ACIS data.

Table 2: Balance of Judge Assignment across Controls

|  | Actual MHT | Judge Propensity |
|---|---|---|
| Age | -0.001*** | 0.000 |
|  | (0.000) | (0.000) |
| Female | 0.008*** | 0.000 |
|  | (0.001) | (0.000) |
| Midwest | 0.031*** | 0.001 |
|  | (0.009) | (0.002) |
| South | 0.007 | -0.001 |
|  | (0.005) | (0.004) |
| West | -0.004 | 0.001 |
|  | (0.012) | (0.002) |
| Black | -0.020*** | -0.000 |
|  | (0.001) | (0.000) |
| Hispanic | -0.008*** | -0.000 |
|  | (0.002) | (0.000) |
| Domestic Violence or Sex Offense | 0.051*** | 0.003*** |
|  | (0.003) | (0.001) |
| First Time in System | 0.009*** | 0.000 |
|  | (0.001) | (0.000) |
| Prior Arrest in the Last Year | -0.002*** | -0.000 |
|  | (0.001) | (0.000) |
| Private Attorney | 0.017*** | 0.001*** |
|  | (0.001) | (0.000) |
| Joint F-Stat | 404.9 | 2.0 |
| P-value | 0.000 | 0.001 |
| Observations | 727,184 | 727,184 |

Note: This table presents results from regressions using either the judge propensity to assign mental health treatment or the mandate to seek mental health treatment as the dependent variable. I regress these on the same vector of controls used in the main specification (age cubic, sex, region of United States, race/ethnicity, first time offender, sex offender, offense groups, prior appearances, attorney) and the court-time fixed effects. I report the F-statistic and associated p-value from a test of joint significance of the controls. The F-statistic is much larger for predicting mental health treatment than for predicting judge propensity toward mental health treatment. Source: 1994-2009 North Carolina ACIS data. Standard errors, clustered at the court-time level, in parentheses. * $p < 0.1$, ** $p < 0.05$, *** $p < 0.01$.

Table 3: First Stage

|  | Mandated Mental Health Treatment | | | |
|---|---|---|---|---|
| MHT Propensity | 0.024*** | 0.024*** | 0.022*** | |
|  | (0.001) | (0.001) | (0.001) | |
| SUDT Propensity |  | 0.001** | 0.001** | |
|  |  | (0.001) | (0.001) | |
| Age at Date of Judgement Issuance |  |  | -0.000 | -0.000 |
|  |  |  | (0.001) | (0.001) |
| Female |  |  | 0.006*** | 0.006*** |
|  |  |  | (0.002) | (0.002) |
| Midwest |  |  | 0.050 | 0.048 |
|  |  |  | (0.031) | (0.031) |
| South |  |  | -0.023 | -0.023 |
|  |  |  | (0.018) | (0.018) |
| West |  |  | -0.057 | -0.061* |
|  |  |  | (0.036) | (0.036) |
| Hispanic |  |  | 0.008 | 0.010 |
|  |  |  | (0.007) | (0.008) |
| Black |  |  | -0.041*** | -0.041*** |
|  |  |  | (0.002) | (0.002) |
| Domestic Violence or Sex Offense |  |  | 0.054*** | 0.062*** |
|  |  |  | (0.007) | (0.007) |
| First Time Offender |  |  | 0.023*** | 0.024*** |
|  |  |  | (0.002) | (0.002) |
| Prior Arrest in Last Year |  |  | -0.001 | -0.000 |
|  |  |  | (0.002) | (0.002) |
| Public Attorney |  |  | -0.001 | -0.002 |
|  |  |  | (0.002) | (0.003) |
| Private Attorney |  |  | 0.033*** | 0.033*** |
|  |  |  | (0.003) | (0.003) |
| Observations | 727,184 | 727,184 | 727,184 | 727,184 |
| Adj. $R^2$ | 0.063 | 0.063 | 0.082 | 0.071 |

Note: This table presents results for the first-stage of the main specification. These results come from regressing the mandate to seek mental health treatment on the standardized judge propensity instruments for mental health treatment (MHT) and substance use disorder treatment (SUDT) and individual characteristics. The controls included are race, ethnicity, sex, age, region, first-time offender, attorney, offense types, whether the offender had a prior offense in the past year, and county-level descriptors: per-capita psychologists, emergency room visits related to mental illness, Medicaid coverage, uninsurance rate, poverty rate, violent crime rate, and proportion with at least a bachelor's degree. Source: 1994-2009 North Carolina ACIS data. Standard errors, clustered at the court-time level, in parentheses. * p < 0.1, ** p < 0.05, *** p < 0.01.

Table 4: Effect of Mental Health Treatment on Recidivism

|  | OLS | | IV | | | |
|---|---|---|---|---|---|---|
| Assigned MHT | −0.102*** | −0.047*** | −0.162*** | −0.148*** | −0.122*** | −0.121*** |
|  | (0.003) | (0.003) | (0.045) | (0.045) | (0.044) | (0.044) |
| 1st Stage F-Stat |  |  | 1024 | 1008 | 990 | 970 |
| Mean of Outcome | 0.342 | 0.342 | 0.342 | 0.342 | 0.342 | 0.342 |
| Court-time FEs | X | X | X | X | X | X |
| Demographic |  | X |  | X | X | X |
| Criminal History |  | X |  |  | X | X |
| County |  | X |  |  |  | X |
| Observations | 727,184 | 727,184 | 727,184 | 727,184 | 727,184 | 727,184 |
| Adj. $R^2$ | 0.032 | 0.095 | 0.022 | 0.039 | 0.064 | 0.066 |

Note: This table presents results from OLS and IV regressions of three-year recidivism on being mandated to seek mental health treatment. All specifications condition on the judge's tendency toward substance use disorder treatment as well as on court-time effects. The third column has no controls other than the court-time fixed effects and offense types; the fourth includes controls for race, ethnicity, sex, age, and region; the fifth adds controls for first-time offender, attorney, offense types, and whether the offender had a prior offense in the past year; and the sixth column adds county-level descriptors: per-capita psychologists, emergency room visits related to mental illness, Medicaid coverage, uninsurance rate, poverty rate, violent crime rate, and proportion with at least a bachelor's degree. Source: 1994-2009 North Carolina ACIS data. Standard errors, clustered at the court-time level, in parentheses. * $p < 0.1$, ** $p < 0.05$, *** $p < 0.01$.

Table 5: Effect of Mental Health Treatment on Recidivism, Excluding Probation Violations

|  | Fail Probation | | Recidivate, Not Counting Fail | |
|---|---|---|---|---|
|  | OLS | IV | OLS | IV |
| Assigned MHT | 0.008*** | 0.026 | −0.050*** | −0.129*** |
|  | (0.003) | (0.036) | (0.003) | (0.045) |
| 1st Stage F-Stat |  | 970 |  | 970 |
| Mean of Outcome | 0.056 | 0.056 | 0.325 | 0.325 |
| Observations | 727,184 | 727,184 | 727,184 | 727,184 |
| Adj. $R^2$ | 0.091 | 0.062 | 0.097 | 0.068 |

Note: This table presents results from OLS and IV regressions of either failing probation or recidivism (defined so that probation violations do not count as committing a future crime) on being mandated to seek mental health treatment. All specifications condition on the judge's tendency toward substance use disorder treatment as well as on court-time effects. The included controls are race, ethnicity, sex, age cubic, first-time offender, region of the United States, shift, day of week, month, year, attorney type, offense types, whether the offender had a prior offense in the past year, and county-level descriptors. Source: 1994-2009 North Carolina ACIS data. Standard errors, clustered at the court-time level, in parentheses. * $p < 0.1$, ** $p < 0.05$, *** $p < 0.01$.

Table 6: Effect of Mental Health Treatment within Charged Felony Type

|  | Misdemeanor | | Felony | |
|---|---|---|---|---|
|  | OLS | IV | OLS | IV |
| Assigned MHT | −0.071*** | −0.101** | −0.050*** | −0.177** |
|  | (0.003) | (0.048) | (0.005) | (0.071) |
| 1st Stage F-Stat |  | 769 |  | 589 |
| Mean of Outcome | 0.31 | 0.31 | 0.37 | 0.37 |
| Prop. Assigned MHT | 0.07 | 0.07 | 0.03 | 0.03 |
| Observations | 305,420 | 305,386 | 421,764 | 421,425 |
| Adj. $R^2$ | 0.080 | 0.032 | 0.067 | 0.041 |

Note: This table presents results from OLS and IV regressions of three-year recidivism on mental health treatment, estimated separately by misdemeanor and felony offenses. All specifications condition on the judge's tendency toward substance use disorder treatment as well as on court-time effects. The included controls are race, ethnicity, sex, age cubic, first-time offender, region of the United States, shift, day of week, month, year, attorney type, offense types, whether the offender had a prior offense in the past year, and county-level descriptors. Source: 1994-2009 North Carolina ACIS data. Standard errors, clustered at the court-time level, in parentheses. * $p < 0.1$, ** $p < 0.05$, *** $p < 0.01$.

Table 7: Effect of Mental Health Treatment Among Offense Groups

|  | Violent/Property | Financial/Fraud | Drugs/Alcohol | Traffic/Public Order | Miscellaneous |
|---|---|---|---|---|---|
| Assigned MHT | −0.846** | −0.972* | −0.175** | 0.025 | 0.069 |
|  | (0.353) | (0.691) | (0.084) | (0.098) | (0.174) |
| 1st Stage F-Stat | 116 | 63 | 285 | 252 | 70 |
| Mean of Outcome | 0.42 | 0.42 | 0.29 | 0.29 | 0.27 |
| Prop. Assigned MHT | 0.03 | 0.02 | 0.05 | 0.07 | 0.03 |
| Observations | 195,659 | 115,313 | 129,436 | 264,023 | 21,154 |
| Adj. $R^2$ | −0.035 | −0.132 | 0.022 | 0.009 | 0.025 |

Note: This table presents results from IV regressions of three-year recidivism on mental health treatment, estimated by five offense groups which are mutually exclusive and exhaustive. All specifications condition on the judge's tendency toward substance use disorder treatment as well as on court-time effects. The included controls are race, ethnicity, sex, age cubic, first-time offender, region of the United States, shift, day of week, month, year, attorney type, offense types, whether the offender had a prior offense in the past year, and county-level descriptors. Source: 1994-2009 North Carolina ACIS data. Standard errors, clustered at the court-time level, in parentheses. * $p < 0.1$, ** $p < 0.05$, *** $p < 0.01$.

Table 8: Effect of Mental Health Treatment on Number of Crimes

|  | OLS | | IV | |
|---|---|---|---|---|
| Assigned MHT | $-0.191$*** | $-0.084$*** | $-0.351$*** | $-0.214$** |
|  | (0.006) | (0.005) | (0.093) | (0.083) |
| 1st Stage F-Stat |  |  | 1149 | 1126 |
| Mean of Outcome | 0.539 | 0.539 | 0.539 | 0.539 |
| Controls |  | X |  | X |
| Observations | 727,184 | 727,184 | 727,184 | 727,184 |
| Adj. $R^2$ | 0.032 | 0.098 | 0.001 | 0.069 |

Note: This table presents results from OLS and IV regressions of the number of future crimes on mental health treatment. All specifications condition on the judge's tendency toward substance use disorder treatment as well as on court-time effects. The included controls are race, ethnicity, sex, age cubic, first-time offender, region of the United States, shift, day of week, month, year, attorney type, offense types, whether the offender had a prior offense in the past year, and county-level descriptors. Source: 1994-2009 North Carolina ACIS data. Standard errors, clustered at the court-time level, in parentheses. * $p < 0.1$, ** $p < 0.05$, *** $p < 0.01$.

Table 9: Effect of Mental Health Treatment on Future Sentence

|  | Unconditional Sentence Length | | Conditional Sentence Length | | Unconditional Active Sentence | | Conditional Active Sentence | |
|---|---|---|---|---|---|---|---|---|
|  | OLS | IV | OLS | IV | OLS | IV | OLS | IV |
| Assigned MHT | −17.971*** | −55.366* | −30.155*** | −110.984 | −0.026*** | −0.062** | −0.053*** | −0.091 |
|  | (1.829) | (28.528) | (7.004) | (105.882) | (0.002) | (0.031) | (0.006) | (0.094) |
| 1st Stage F-Stat |  | 970 |  | 352 |  | 970 |  | 352 |
| Mean of Outcome | 96.693 | 96.693 | 282.352 | 282.352 | 0.142 | 0.142 | 0.416 | 0.416 |
| Observations | 727,184 | 727,184 | 249,029 | 248,797 | 727,184 | 727,184 | 249,029 | 248,797 |
| Adj. $R^2$ | 0.032 | 0.017 | 0.060 | 0.023 | 0.071 | 0.045 | 0.109 | 0.046 |

Note: This table presents results from OLS and IV regressions of the qualities of the future sentence (average sentence length measured in days and whether the sentence is an active prison sentence) on mental health treatment. All specifications condition on the judge's tendency toward substance use disorder treatment as well as on court-time effects. The included controls are race, ethnicity, sex, age cubic, first-time offender, region of the United States, shift, day of week, month, year, attorney type, offense types, whether the offender had a prior offense in the past year, and county-level descriptors. Source: 1994-2009 North Carolina ACIS data. Standard errors, clustered at the court-time level, in parentheses. * $p < 0.1$, ** $p < 0.05$, *** $p < 0.01$.

Table 10: Effect of Mental Health Treatment on Type of Offense Committed

|  | Violent/Property | | Financial/Fraud | | Drugs/Alcohol | | Traffic/Public Order | | Miscellaneous | |
| --- | --- | --- | --- | --- | --- | --- | --- | --- | --- | --- |
|  | OLS | IV | OLS | IV | OLS | IV | OLS | IV | OLS | IV |
| Assigned MHT | −0.012*** | −0.111 | −0.017*** | −0.175** | −0.012*** | −0.040 | −0.006*** | −0.056 | −0.000 | 0.000 |
|  | (0.002) | (0.074) | (0.002) | (0.083) | (0.001) | (0.062) | (0.002) | (0.082) | (0.000) | (0.014) |
| 1st Stage F-Stat |  | 348 |  | 348 |  | 348 |  | 348 |  | 348 |
| Mean of Outcome | 0.10 | 0.10 | 0.09 | 0.09 | 0.07 | 0.07 | 0.08 | 0.08 | 0.00 | 0.00 |
| Observations | 727,184 | 727,184 | 727,184 | 727,184 | 727,184 | 727,184 | 727,184 | 727,184 | 727,184 | 727,184 |
| Adj. $R^2$ | 0.080 | 0.052 | 0.076 | 0.037 | 0.034 | 0.017 | 0.042 | 0.012 | 0.015 | 0.003 |

Note: This table presents results from OLS and IV regressions of the type of future crime on mental health treatment. These results are *not* conditional on having committed a future crime. All specifications condition on the judge's tendency toward substance use disorder treatment as well as on court-time effects. The included controls are race, ethnicity, sex, age cubic, first-time offender, region of the United States, shift, day of week, month, year, attorney type, offense types, whether the offender had a prior offense in the past year, and county-level descriptors. Source: 1994-2009 North Carolina ACIS data. Standard errors, clustered at the court-time level, in parentheses. * $p < 0.1$, ** $p < 0.05$, *** $p < 0.01$.

Table 11: Effect of Mental Health Treatment on Severity of Offense Committed

|  | Misdemeanor | | Felony | |
| --- | --- | --- | --- | --- |
|  | OLS | IV | OLS | IV |
| Assigned MHT | −0.008*** | 0.003 | −0.039*** | −0.120*** |
|  | (0.002) | (0.024) | (0.003) | (0.043) |
| 1st Stage F-Stat |  | 970 |  | 970 |
| Mean of Outcome | 0.07 | 0.07 | 0.27 | 0.27 |
| Observations | 727,184 | 727,184 | 727,184 | 727,184 |
| Adj. $R^2$ | 0.077 | 0.019 | 0.112 | 0.067 |

Note: This table presents results from OLS and IV regressions of whether the future crime was a misdemeanor or a felony on mental health treatment. These results are *not* conditional on having committed a future crime. All specifications condition on the judge's tendency toward substance use disorder treatment as well as on court-time effects. The included controls are race, ethnicity, sex, age cubic, first-time offender, region of the United States, shift, day of week, month, year, attorney type, offense types, whether the offender had a prior offense in the past year, and county-level descriptors. Source: 1994-2009 North Carolina ACIS data. Standard errors, clustered at the court-time level, in parentheses. * $p < 0.1$, ** $p < 0.05$, *** $p < 0.01$.

Table 12: Effect of Mental Health Treatment on Recidivism, Without Controlling for Drug Propensity

|  | OLS | | IV | |
|---|---|---|---|---|
| Assigned MHT | −0.102*** | −0.047*** | −0.175*** | −0.141*** |
|  | (0.003) | (0.003) | (0.042) | (0.041) |
| 1st Stage F-Stat |  |  | 1169 | 1126 |
| Mean of Outcome | 0.342 | 0.342 | 0.342 | 0.342 |
| Controls |  | X |  | X |
| Observations | 727,184 | 727,184 | 727,184 | 727,184 |
| Adj. $R^2$ | 0.032 | 0.095 | 0.021 | 0.065 |

Note: This table presents results from OLS and IV regressions of three-year recidivism on mental health treatment. The specification is identical to the main specification except that the judge propensity to assign substance use disorder treatment is not included as a control. All specifications condition on court-time effects. The included controls are race, ethnicity, sex, age cubic, first-time offender, region of the United States, shift, day of week, month, year, attorney type, offense types, whether the offender had a prior offense in the past year, and county-level descriptors. Source: 1994-2009 North Carolina ACIS data. Standard errors, clustered at the court-time level, in parentheses. * $p < 0.1$, ** $p < 0.05$, *** $p < 0.01$.

Figure 1: Rate of Recidivism over Time Since Trial

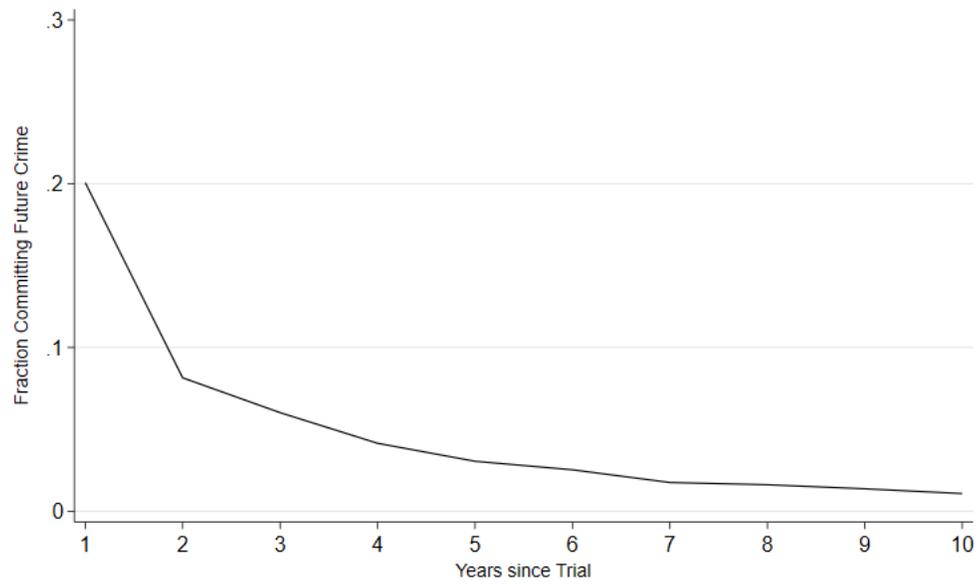

Note: This figure shows how many probationers are convicted of a new crime in each year since the trial. The highest risk of recidivism by far is in the first year following a conviction. Source: 1994-2009 North Carolina ACIS data.

Figure 2: Distribution of Offense Classes Among Probationers

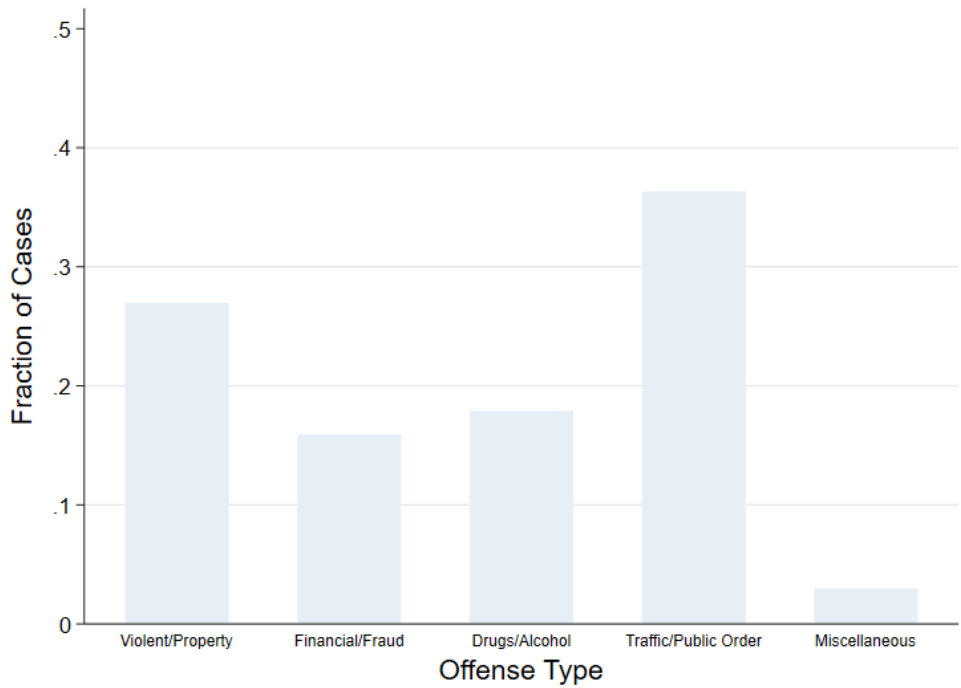

Note: This figure shows how often each offense type was committed by probationers. For each case, the defining offense is chosen to be the most severe offense from among those charged. Source: 1994-2009 North Carolina ACIS data.

Figure 3: Frequency of Mental Health Treatment Among Offense Classes

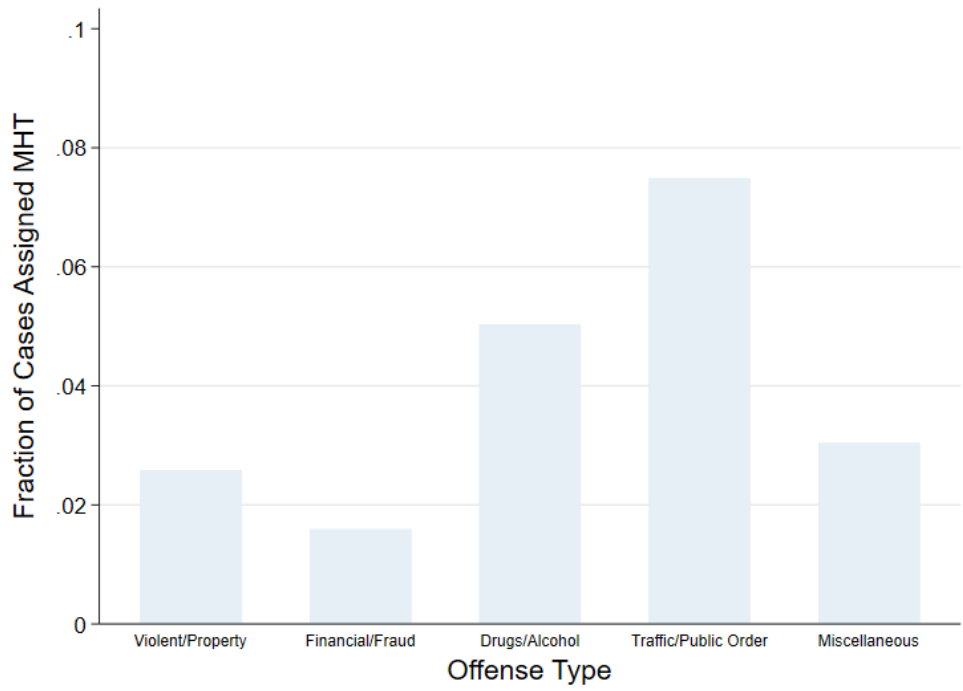

Note: This figure shows how often mental health treatment was assigned as a term of probation among different offense types. For each case, the defining offense is chosen to be the most severe offense from among those charged. Source: 1994-2009 North Carolina ACIS data.

Figure 4: Judges with a Higher Propensity Measure Assign More Mental Health Treatment

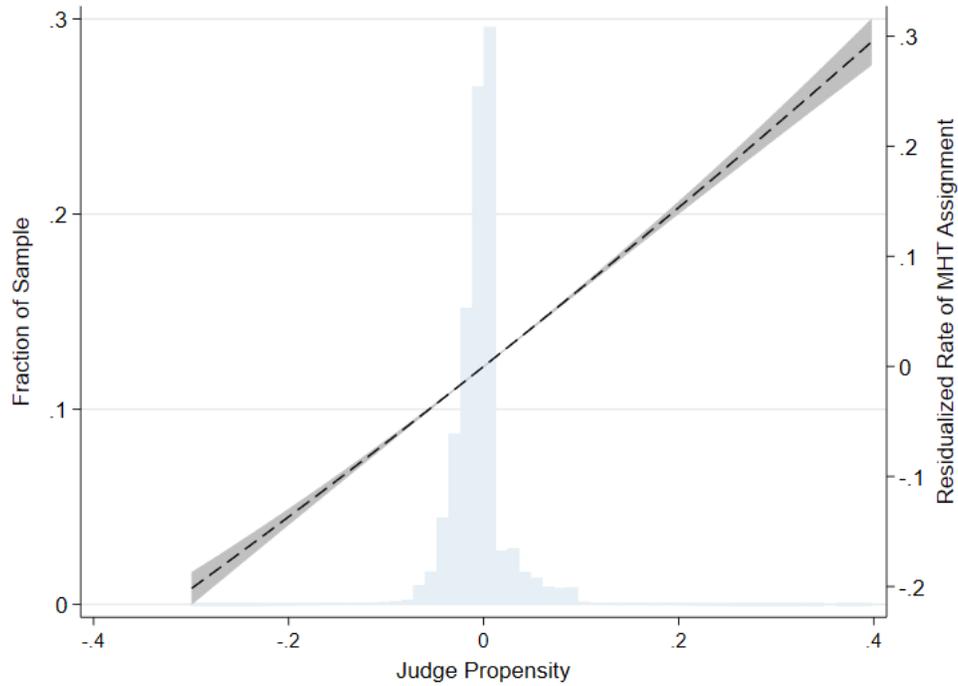

Note: On the left axis, this figure plots a histogram of the instrument, which is a measure of the judge tendency toward mandating mental health treatment. For the purposes of this graph I remove right-tail outlier judges (.7% of the sample). Positive values represent those judges who have a higher tendency toward mental health treatment. On the right axis, this figure plots mental health treatment, residualized on controls, against the measure of judge propensity. This provides a graphical representation of the first stage: judges with a higher propensity are more likely to mandate mental health treatment. Source: 1994-2009 North Carolina ACIS data.

Figure 5: Proportion of Time Judges Assign Mental Health Treatment

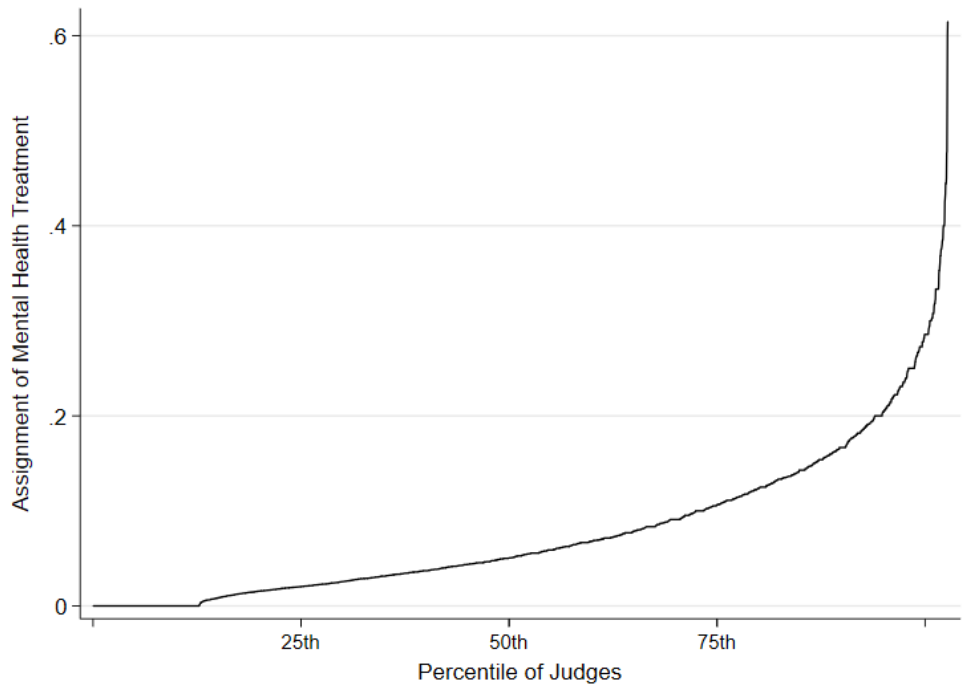

Note: This figure plots each judge's rate of mental health treatment assignment, ordered from those judges who are least likely to those who are most likely to assign mental health treatment. Source: 1994-2009 North Carolina ACIS data.

Figure 6: Effect of Mental Health Treatment on Recidivism, by Individual Characteristics

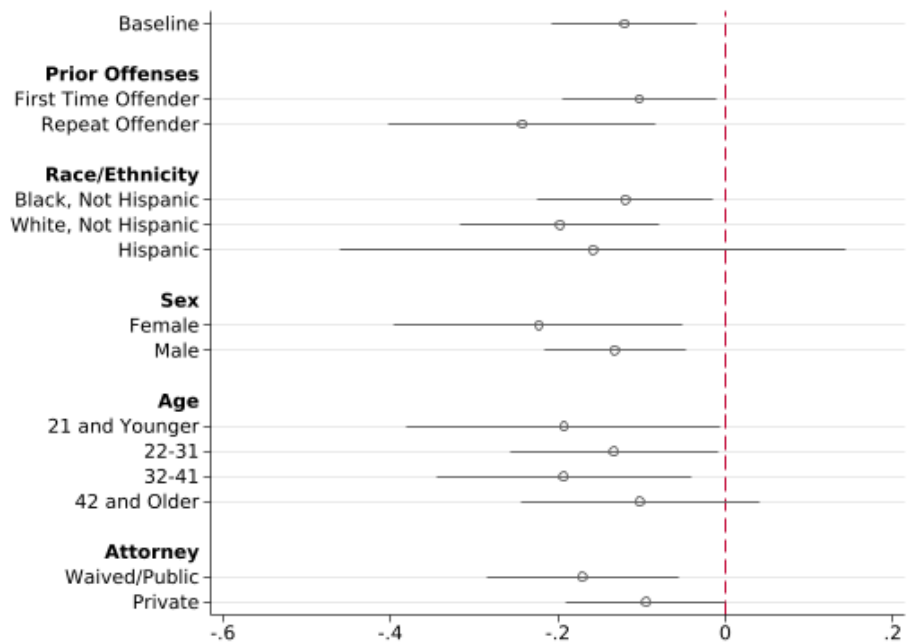

Note: This figure plots the percentage point decline in three-year recidivism due to mandated mental health treatment, estimated from the main IV regression with full controls. I estimate the effect separately for groups defined by five characteristics: first-time offender, race/ethnicity, sex, age, and attorney type. All specifications condition on the judge's tendency toward substance use disorder treatment as well as on court-time effects. The included controls are race, ethnicity, sex, age cubic, first-time offender, region of the United States, shift, day of week, month, year, attorney type, offense types, whether the offender had a prior offense in the past year, and county-level descriptors. Source: 1994-2009 North Carolina ACIS data.

Figure 7: Effect of Mental Health Treatment on Failing Probation, by Individual Characteristics

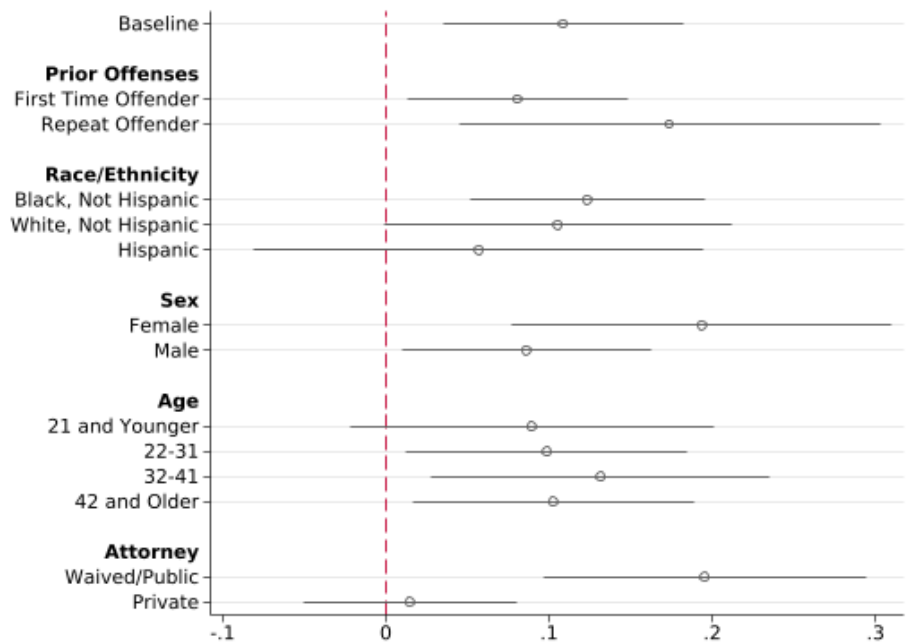

Note: This graph plots the percentage point effect on violating probation due to mental health treatment, estimated from the main IV regression with full controls. I estimate the effect separately for groups defined by five characteristics: first-time offender, race/ethnicity, sex, age, and attorney type. All specifications condition on the judge's tendency toward substance use disorder treatment as well as on court-time effects. The included controls are race, ethnicity, sex, age cubic, first-time offender, region of the United States, shift, day of week, month, year, attorney type, offense types, whether the offender had a prior offense in the past year, and county-level descriptors. Source: 1994-2009 North Carolina ACIS data.

Figure 8: The Effect of Mental Health Treatment on Recidivism over Time

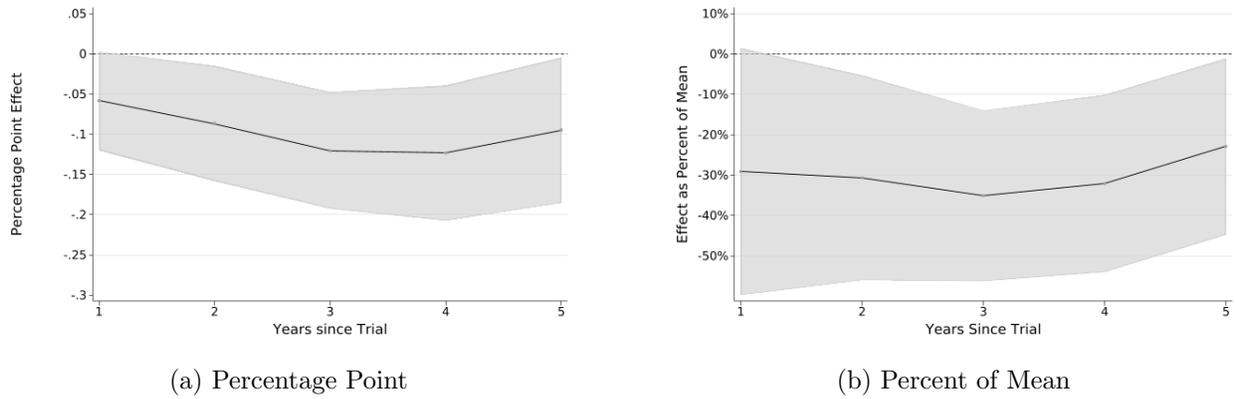

(a) Percentage Point

(b) Percent of Mean

Note: This figure plots the percentage point (left) and the percent (right) decline in recidivism due to mental health treatment, estimated from the main IV regression with full controls. Recidivism is defined first looking only one year out from the trial, then two years, and so on up to five years. All specifications condition on the judge's tendency toward substance use disorder treatment as well as on court-time effects. The included controls are race, ethnicity, sex, age cubic, first-time offender, region of the United States, shift, day of week, month, year, attorney type, offense types, whether the offender had a prior offense in the past year, and county-level descriptors. Source: 1994-2009 North Carolina ACIS data.

Figure 9: The Effect of Mental Health Treatment on Recidivism in Each Year Since Trial

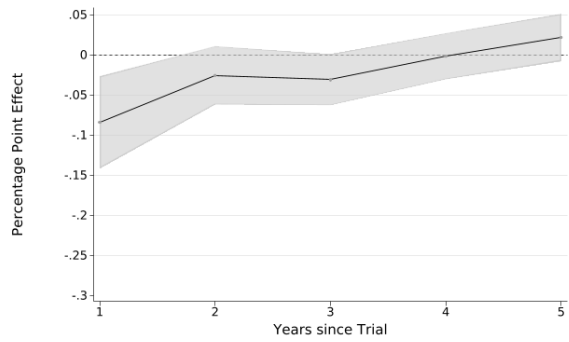

(a) Percentage Point

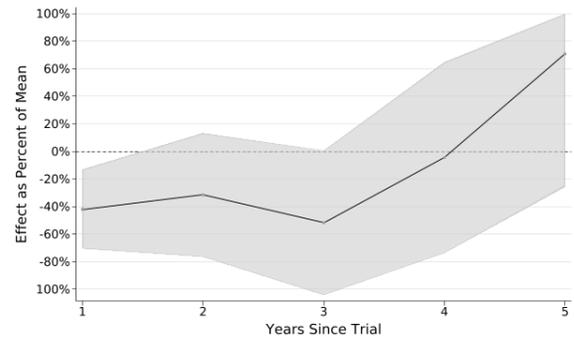

(b) Percent of Mean

Note: This figure plots the percentage point (left) and the percent (right) decline in recidivism due to mental health treatment, estimated from the main IV regression with full controls. Recidivism is defined first looking only within one year out from the trial, then between one and two years, then between two and three years, and so on up to five years. All specifications condition on the judge's tendency toward substance use disorder treatment as well as on court-time effects. The included controls are race, ethnicity, sex, age cubic, first-time offender, region of the United States, shift, day of week, month, year, attorney type, offense types, whether the offender had a prior offense in the past year, and county-level descriptors. Source: 1994-2009 North Carolina ACIS data.

Figure 10: The Effect of Mental Health Treatment on Recidivism over Time, by Type of Attorney

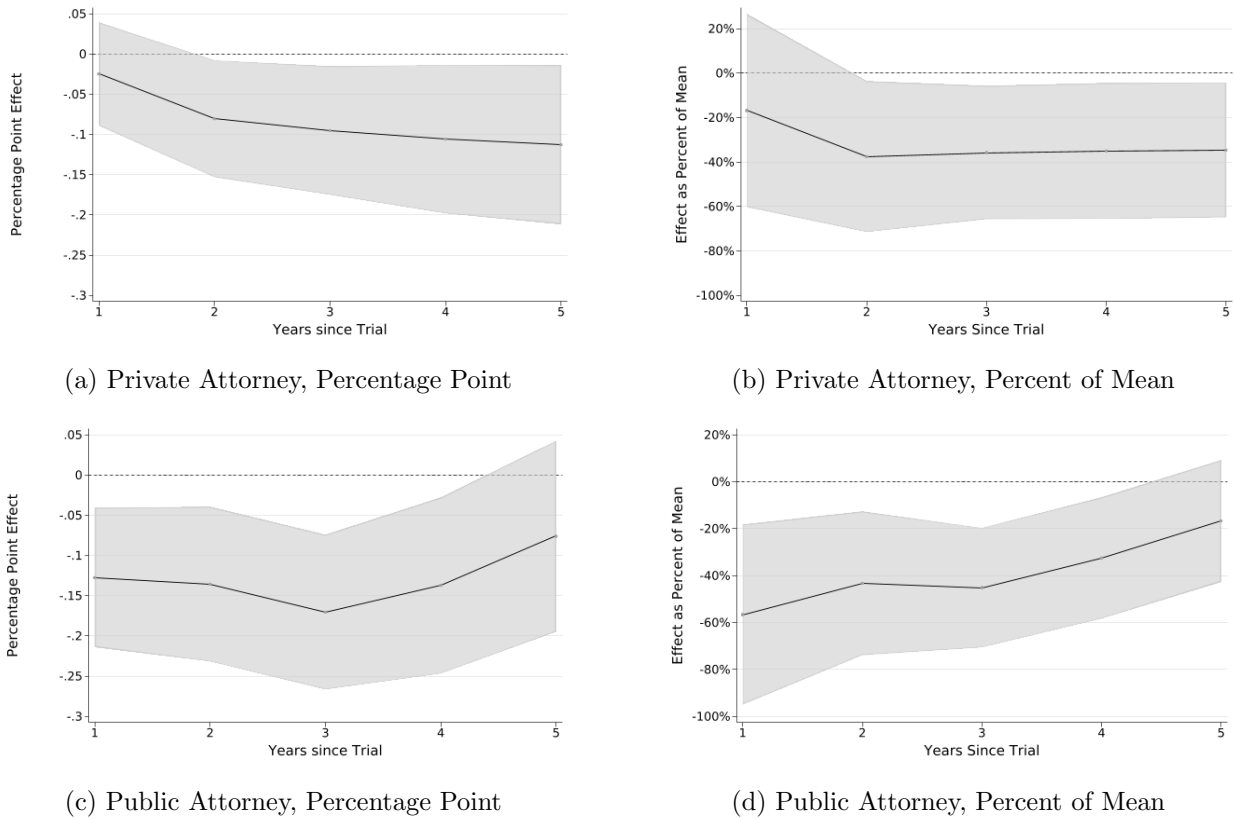

(a) Private Attorney, Percentage Point

(b) Private Attorney, Percent of Mean

(c) Public Attorney, Percentage Point

(d) Public Attorney, Percent of Mean

Note: This figure plots the percentage point (left) and percent (right) decline in recidivism due to mental health treatment, estimated from the main IV regression with full controls. The effect is estimated separately among individuals with private attorneys (top row) and those with public attorneys (bottom row). Recidivism is defined first looking only one year out from the trial, then two years, and so on up to five years. All specifications condition on the judge's tendency toward substance use disorder treatment as well as on court-time effects. The included controls are race, ethnicity, sex, age cubic, first-time offender, region of the United States, shift, day of week, month, year, attorney type, offense types, whether the offender had a prior offense in the past year, and county-level descriptors. Source: 1994-2009 North Carolina ACIS data.

Figure 11: Costs and Benefits of Mental Health Treatment During Probation

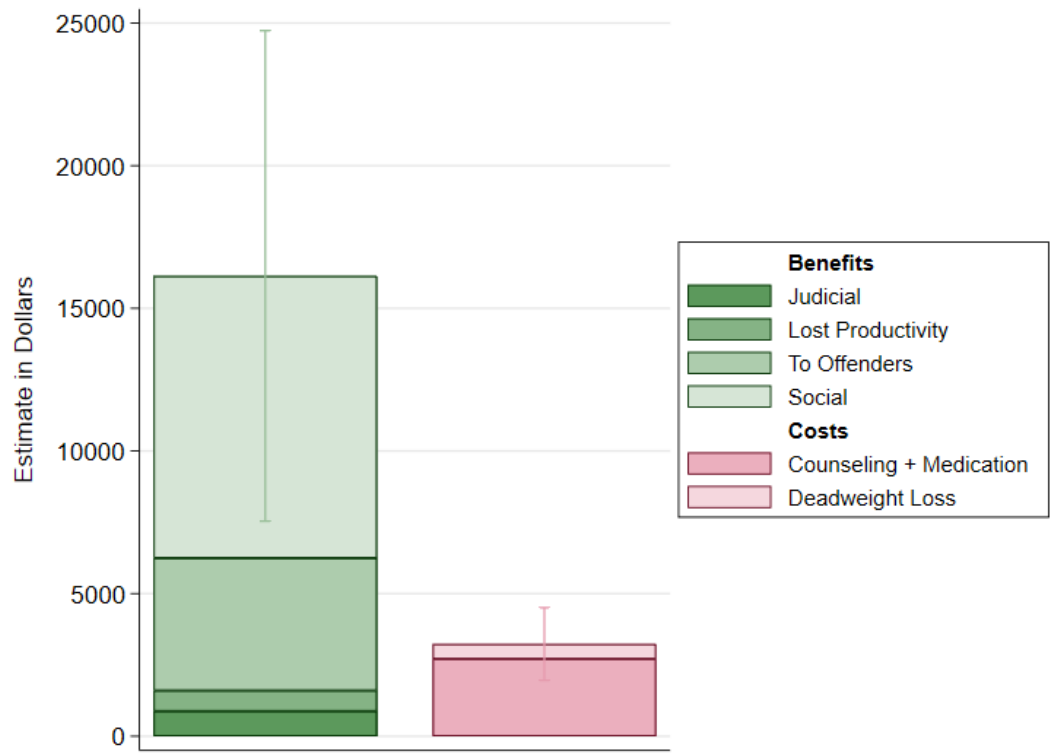

Note: This figure plots the costs to the government of providing mental health treatment (in red) alongside the monetary value of the benefits of mental health treatment through the decreased incidence of crime (in green). The large confidence intervals incorporate both sampling variation from the empirical estimates as well variation from the cost-of-crime literature as to the monetary valuation of different types of crimes. Comparing the size of the bars suggests that mandated mental health treatment is about 5 times as beneficial as it is costly.

# A  Appendix: Tables and Figures

Table A1: Offender Characteristics Before Limiting to Probation Cases

|                            | All              | Limit to Probation |
|----------------------------|------------------|--------------------|
| Age                        | 30.94<br>(10.41) | 30.80<br>(10.37)   |
| Female                     | 0.18<br>(0.38)   | 0.20<br>(0.40)     |
| Black                      | 0.47<br>(0.50)   | 0.45<br>(0.50)     |
| Hispanic                   | 0.02<br>(0.16)   | 0.03<br>(0.16)     |
| Violent/Property Crime     | 0.34<br>(0.47)   | 0.27<br>(0.44)     |
| Drug/Alcohol Crime         | 0.18<br>(0.38)   | 0.19<br>(0.39)     |
| Financial/Fraud Crime      | 0.13<br>(0.33)   | 0.15<br>(0.35)     |
| Traffic/Public Order Crime | 0.32<br>(0.47)   | 0.36<br>(0.48)     |
| Sex Offender               | 0.01<br>(0.12)   | 0.01<br>(0.11)     |
| First Time Offender        | 0.58<br>(0.49)   | 0.59<br>(0.49)     |
| Private Attorney           | 0.28<br>(0.45)   | 0.30<br>(0.46)     |
| Assigned MH Treatment      | 0.05<br>(0.22)   | 0.05<br>(0.21)     |
| Assigned SUD Treatment     | 0.05<br>(0.21)   | 0.04<br>(0.20)     |
| Recidivism within 3 Years  | 0.33<br>(0.47)   | 0.34<br>(0.47)     |
| Observations               | 1,439,406        | 1,039,994          |

Note: This table presents summary statistics for the sample before and after the restriction to probation cases is made. The second column is a subset of the sample in the first column. I designate First-Time Offender based on when offenders first appear within the sample period beginning in 1994, which could misclassify offenders who were arrested in years prior to 1994. "SUD" stands for Substance Use Disorder. Source: 1994-2009 North Carolina ACIS data.

Table A2: Effect of Mental Health Treatment on Recidivism, by Years Since Trial

|  | 1 Year | 2 Years | 3 Years | 4 Years | 5 Years |
|---|---|---|---|---|---|
| Assigned MHT | −0.058 | −0.087** | −0.120*** | −0.123** | −0.095* |
|  | (0.037) | (0.044) | (0.044) | (0.051) | (0.055) |
| 1st Stage F-Stat | 970 | 970 | 970 | 829 | 690 |
| Mean of Dependent Var | 0.20 | 0.28 | 0.34 | 0.38 | 0.41 |
| Observations | 727,184 | 727,184 | 727,184 | 632,772 | 547,732 |
| Adj. $R^2$ | 0.045 | 0.056 | 0.066 | 0.074 | 0.083 |

Note: This table presents results from OLS and IV regressions of recidivism on being mandated to seek mental health treatment. Recidivism is defined first looking only one year out from the trial, then two years, and so on up to five years. All specifications condition on the judge's tendency toward substance use disorder treatment as well as on court-time effects. The included controls are race, ethnicity, sex, age cubic, first-time offender, region of the United States, shift, day of week, month, year, attorney type, offense types, whether the offender had a prior offense in the past year, and county-level descriptors. Source: 1994-2009 North Carolina ACIS data. Standard errors, clustered at the court-time level, in parentheses. * $p < 0.1$, ** $p < 0.05$, *** $p < 0.01$.

Table A3: Robustness: Comparison of Different Instrument Cluster Levels

|  | Base | Remove Time Variables | Remove Circuit |
|---|---|---|---|
| Assigned MHT | −0.121*** | −0.139*** | −0.152*** |
|  | (0.044) | (0.053) | (0.055) |
| 1st Stage F-Stat | 970 | 428 | 402 |
| Mean of Outcome | 0.342 | 0.342 | 0.342 |
| Observations | 727,184 | 727,184 | 727,184 |
| Adj. $R^2$ | 0.066 | 0.067 | 0.067 |

Note: This table presents results from OLS and IV regressions of three-year recidivism on being mandated to seek mental health treatment. All specifications condition on the judge's tendency toward substance use disorder treatment as well as on court-time effects, but the columns vary in how the court-time effects are defined. Column 1 presents the baseline specification. Column 2 removes time variables, using only the district or circuit-district to form the groups. Column 3 removes circuit as well. The included controls are race, ethnicity, sex, age cubic, first-time offender, region of the United States, shift, day of week, month, year, attorney type, offense types, whether the offender had a prior offense in the past year, and county-level descriptors. Source: 1994-2009 North Carolina ACIS data. Standard errors, clustered at the court-time level, in parentheses. * $p < 0.1$, ** $p < 0.05$, *** $p < 0.01$.

Table A4: Robustness: Comparison of Different Instrument Time Horizons

|                   | Base        | First Year | Three Years | By Year     | Omit Cluster |
|-------------------|-------------|------------|-------------|-------------|--------------|
| Assigned MHT      | −0.121***   | −0.052     | −0.118***   | −0.122***   | −0.127***    |
|                   | (0.044)     | (0.115)    | (0.039)     | (0.047)     | (0.047)      |
| 1st Stage F-Stat  | 970         | 104        | 937         | 574         | 732          |
| Mean of Outcome   | 0.342       | 0.339      | 0.342       | 0.343       | 0.342        |
| Observations      | 727,184     | 620,892    | 725,925     | 723,679     | 726,822      |
| Adj. $R^2$        | 0.066       | 0.067      | 0.066       | 0.066       | 0.065        |

Note: This table presents results from OLS and IV regressions of three-year recidivism on being mandated to seek mental health treatment. The columns vary by the time horizons used to create the instrument. Column 1 presents the baseline specification. Column 2 uses only cases from the judge's first year to form the instrument. Column 3 creates the instrument within groups of three years. Column 4 creates the instrument by each year. Column 5 omits cases within the same court-time group. All specifications condition on the judge's tendency toward substance use disorder treatment as well as on court-time effects. The included controls are race, ethnicity, sex, age cubic, first-time offender, region of the United States, shift, day of week, month, year, attorney type, offense types, whether the offender had a prior offense in the past year, and county-level descriptors. Source: 1994-2009 North Carolina ACIS data. Standard errors, clustered at the court-time level, in parentheses. * $p < 0.1$, ** $p < 0.05$, *** $p < 0.01$.

Table A5: Robustness: Full Saturation in Covariates

|  | Regular Controls | Saturated Controls | Saturated Instrument | Saturated Controls & Instrument |
|---|---|---|---|---|
| Assigned MHT | −0.121*** | −0.100** | −0.122*** | −0.092** |
|  | (0.044) | (0.045) | (0.044) | (0.044) |
| 1st Stage F-Stat | 970 | 930 | 220 | 214 |
| Mean of Outcome | 0.342 | 0.342 | 0.342 | 0.342 |
| Observations | 727,184 | 713,730 | 727,184 | 713,730 |
| Adj. $R^2$ | 0.066 | 0.054 | 0.066 | 0.054 |

Note: This table presents results from OLS and IV regressions of three-year recidivism on being mandated to seek mental health treatment. Columns 2-4 fully saturate the model with single, double, and triple combinations of the controls and the instruments. All specifications condition on the judge's tendency toward substance use disorder treatment as well as on court-time effects. The included controls are race, ethnicity, sex, age cubic, first-time offender, region of the United States, shift, day of week, month, year, attorney type, offense types, whether the offender had a prior offense in the past year, and county-level descriptors. Source: 1994-2009 North Carolina ACIS data. Standard errors, clustered at the court-time level, in parentheses. * $p < 0.1$, ** $p < 0.05$, *** $p < 0.01$.

Table A6: Robustness: Convictions

|  | Convicted Sample | | Control for Judge Conviction | |
|---|---|---|---|---|
|  | OLS | IV | OLS | IV |
| Assigned MHT | −0.051*** | −0.138*** | −0.047*** | −0.142*** |
|  | (0.003) | (0.046) | (0.003) | (0.044) |
| 1st Stage F-Stat |  | 884 |  | 961 |
| Mean of Outcome | 0.352 | 0.352 | 0.342 | 0.342 |
| Observations | 644,915 | 644,915 | 727,204 | 727,204 |
| $R^2$ | 0.098 | 0.064 | 0.095 | 0.065 |

Note: This table presents results from OLS and IV regressions of three-year recidivism on being mandated to seek mental health treatment. The first two columns limit the sample to only those cases which resulted in a conviction. The second two columns use the main analysis sample, but control for the judge's tendency toward conviction. All specifications condition on the judge's tendency toward substance use disorder treatment as well as on court-time effects. The included controls are race, ethnicity, sex, age cubic, first-time offender, region of the United States, shift, day of week, month, year, attorney type, offense types, whether the offender had a prior offense in the past year, and county-level descriptors. Source: 1994-2009 North Carolina ACIS data. Standard errors, clustered at the court-time level, in parentheses. * $p < 0.1$, ** $p < 0.05$, *** $p < 0.01$.

Table A7: Robustness: Definitions of Mental Health Treatment

|  | Less Restrictive | Base | More Restrictive | Most Restrictive |
|---|---|---|---|---|
| Assigned MHT | −0.152*** | −0.141*** | −0.134*** | −0.165*** |
|  | (0.044) | (0.041) | (0.047) | (0.059) |
| 1st Stage F-Stat | 997 | 1126 | 851 | 762 |
| Mean of Outcome | 0.342 | 0.342 | 0.342 | 0.342 |
| Observations | 727,184 | 727,184 | 727,184 | 727,184 |
| Adj. $R^2$ | 0.065 | 0.065 | 0.065 | 0.065 |

Note: This table presents results from OLS and IV regressions of three-year recidivism on being mandated to seek mental health treatment. The columns vary by which key words were included when defining mental health treatment. All specifications condition on the judge's tendency toward substance use disorder treatment as well as on court-time effects. The included controls are race, ethnicity, sex, age cubic, first-time offender, region of the United States, shift, day of week, month, year, attorney type, offense types, whether the offender had a prior offense in the past year, and county-level descriptors. Source: 1994-2009 North Carolina ACIS data. Standard errors, clustered at the court-time level, in parentheses. * $p < 0.1$, ** $p < 0.05$, *** $p < 0.01$.

Table A8: Robustness: Definitions of SUD (Substance Use Disorder) treatment

|  | Least Restrictive | Less Restrictive | Base | Most Restrictive |
|---|---|---|---|---|
| Assigned MHT | −0.096** | −0.093** | −0.120*** | −0.127*** |
|  | (0.045) | (0.045) | (0.044) | (0.041) |
| 1st Stage F-Stat | 984 | 927 | 970 | 1090 |
| Mean of Outcome | 0.342 | 0.342 | 0.342 | 0.342 |
| Observations | 727,184 | 727,184 | 727,184 | 727,184 |
| Adj. $R^2$ | 0.066 | 0.066 | 0.066 | 0.066 |

Note: This table presents results from OLS and IV regressions of three-year recidivism on being mandated to seek mental health treatment. The columns vary by which key words were included when defining substance use disorder treatment. All specifications condition on the judge's tendency toward substance use disorder treatment as well as on court-time effects. The included controls are race, ethnicity, sex, age cubic, first-time offender, region of the United States, shift, day of week, month, year, attorney type, offense types, whether the offender had a prior offense in the past year, and county-level descriptors. Source: 1994-2009 North Carolina ACIS data. Standard errors, clustered at the court-time level, in parentheses. * $p < 0.1$, ** $p < 0.05$, *** $p < 0.01$.

Table A9: Combining Treatment: Either Mental Health or SUD Treatment

|  | OLS | | IV | |
|---|---|---|---|---|
| MHT or SUDT | −0.093*** | −0.045*** | −0.135*** | −0.129*** |
|  | (0.003) | (0.002) | (0.033) | (0.032) |
| 1st Stage F-Stat |  |  | 1470 | 1345 |
| Mean of Outcome | 0.342 | 0.342 | 0.342 | 0.342 |
| Controls |  | X |  | X |
| Observations | 727,184 | 727,184 | 727,184 | 727,184 |
| Adj. $R^2$ | 0.033 | 0.095 | 0.022 | 0.065 |

Note: This table presents results from OLS and IV regressions of three-year recidivism on being mandated to seek either mental health (MH) or substance use disorder (SUD) treatment. The instrument is the judge's tendency toward mandating either MH or SUD treatment. All specifications condition on court-time effects. The included controls are race, ethnicity, sex, age cubic, first-time offender, region of the United States, shift, day of week, month, year, attorney type, offense types, whether the offender had a prior offense in the past year, and county-level descriptors. Source: 1994-2009 North Carolina ACIS data. Standard errors, clustered at the court-time level, in parentheses. * p < 0.1, ** p < 0.05, *** p < 0.01.

Table A10: Separating Treatment: Disentangling Mental Health from SUD Treatment

|  | MHT | | MHT & SUDT | | SUDT | | All Three | |
| --- | --- | --- | --- | --- | --- | --- | --- | --- |
|  | OLS | IV | OLS | IV | OLS | IV | OLS | IV |
| MHT | −0.052*** | −0.162*** |  |  |  |  | −0.055*** | −0.151*** |
|  | (0.003) | (0.052) |  |  |  |  | (0.003) | (0.054) |
| MHT & SUDT |  |  | −0.034*** | −0.158* |  |  | −0.039*** | −0.010 |
|  |  |  | (0.004) | (0.086) |  |  | (0.004) | (0.090) |
| SUDT |  |  |  |  | −0.037*** | −0.111* | −0.041*** | −0.098 |
|  |  |  |  |  | (0.004) | (0.061) | (0.004) | (0.060) |
| 1st Stage F-Stat |  | 266 |  | 114 |  | 108 |  | 97 |
| Mean of Outcome | 0.342 | 0.342 | 0.342 | 0.342 | 0.342 | 0.342 | 0.342 | 0.342 |
| Observations | 727,184 | 727,060 | 727,184 | 727,060 | 727,184 | 727,060 | 727,184 | 727,060 |
| Adj. $R^2$ | 0.095 | 0.065 | 0.095 | 0.065 | 0.095 | 0.066 | 0.095 | 0.065 |

Note: This table presents results from OLS and IV regressions of three-year recidivism on being mandated to seek four mutually exclusive treatments: mental health (MH) alone, substance use disorder (SUD) treatment alone, MH and SUD treatment together, and neither (which is the left-out category). Each treatment is instrumented with the judge's tendency toward assigning that specific treatment among other offenders in the sample. All specifications condition on court-time effects. The included controls are race, ethnicity, sex, age cubic, first-time offender, region of the United States, shift, day of week, month, year, attorney type, offense types, whether the offender had a prior offense in the past year, and county-level descriptors. Source: 1994-2009 North Carolina ACIS data. Standard errors, clustered at the court-time level, in parentheses. * p < 0.1, ** p < 0.05, *** p < 0.01.

Figure A1: Felony Punishment Classes

| OFFENSE CLASS | PRIOR RECORD LEVEL | | | | | |
|---|---|---|---|---|---|---|
| | I<br>0 Points | II<br>1-4 Points | III<br>5-8 Points | IV<br>9-14 Points | V<br>15-18 Points | VI<br>19+ Points |
| A | Death or Life Without Parole | | | | | |
| B1 | A<br>240 - 300<br>192 - 240<br>144 – 192 | A<br>288 - 360<br>230 - 288<br>173 – 230 | A<br>336 - 420<br>269 - 336<br>202 – 269 | A<br>384 - 480<br>307 - 384<br>230 – 307 | A<br>Life Without Parole<br>346 - 433<br>260 – 346 | A<br>Life Without Parole<br>384 - 480<br>288 - 384 |
| B2 | A<br>157 - 196<br>125 - 157<br>94 - 125 | A<br>189 - 237<br>151 - 189<br>114 - 151 | A<br>220 - 276<br>176 - 220<br>132 - 176 | A<br>251 - 313<br>201 - 251<br>151 - 201 | A<br>282 - 353<br>225 - 282<br>169 - 225 | A<br>313 - 392<br>251 - 313<br>188 - 251 |
| C | A<br>73 – 92<br>58 - 73<br>44 - 58 | A<br>100 – 125<br>80 - 100<br>60 - 80 | A<br>116 – 145<br>93 - 116<br>70 - 93 | A<br>133 - 167<br>107 - 133<br>80 - 107 | A<br>151 - 188<br>121 - 151<br>90 - 121 | A<br>168 - 210<br>135 - 168<br>101 - 135 |
| D | A<br>64 - 80<br>51 - 64<br>38 - 51 | A<br>77 - 95<br>61 - 77<br>46 - 61 | A<br>103 - 129<br>82 - 103<br>61 - 82 | A<br>117 - 146<br>94 - 117<br>71 - 94 | A<br>133 - 167<br>107 - 133<br>80 - 107 | A<br>146 - 183<br>117 - 146<br>88 - 117 |
| E | I/A<br>25 - 31<br>20 - 25<br>15 - 20 | I/A<br>29 - 36<br>23 - 29<br>17 - 23 | A<br>34 - 42<br>27 - 34<br>20 - 27 | A<br>46 - 58<br>37 - 46<br>28 - 37 | A<br>53 - 66<br>42 - 53<br>32 - 42 | A<br>59 - 74<br>47 - 59<br>35 - 47 |
| F | I/A<br>16 - 20<br>13 - 16<br>10 - 13 | I/A<br>19 - 24<br>15 - 19<br>11 - 15 | I/A<br>21 - 26<br>17 - 21<br>13 - 17 | A<br>25 - 31<br>20 - 25<br>15 - 20 | A<br>34 - 42<br>27 - 34<br>20 - 27 | A<br>39 - 49<br>31 - 39<br>23 - 31 |
| G | I/A<br>13 - 16<br>10 - 13<br>8 - 10 | I/A<br>15 - 19<br>12 - 15<br>9 - 12 | I/A<br>16 - 20<br>13 - 16<br>10 - 13 | I/A<br>20 - 25<br>16 - 20<br>12 - 16 | A<br>21 - 26<br>17 - 21<br>13 - 17 | A<br>29 - 36<br>23 - 29<br>17 - 23 |
| H | C/I/A<br>6 - 8<br>5 - 6<br>4 - 5 | I/A<br>8 - 10<br>6 - 8<br>4 - 6 | I/A<br>10 - 12<br>8 - 10<br>6 - 8 | I/A<br>11 - 14<br>9 - 11<br>7 - 9 | I/A<br>15 - 19<br>12 - 15<br>9 - 12 | A<br>20 - 25<br>16 - 20<br>12 - 16 |
| I | C<br>6 - 8<br>4 - 6<br>3 - 4 | C/I<br>6 - 8<br>4 - 6<br>3 - 4 | I<br>6 - 8<br>5 - 6<br>4 - 5 | I/A<br>8 - 10<br>6 - 8<br>4 - 6 | I/A<br>9 - 11<br>7 - 9<br>5 - 7 | I/A<br>10 - 12<br>8 - 10<br>6 - 8 |

Note: This chart shows how the class of the current offense and the number and severity of prior convictions determine different levels of punishment and sentence lengths. A = Active, I = Intermediate, C = Community. The numbers in the cells represent sentence minimums in months. The majority of felony classes involve an active punishment (incarceration). Source: North Carolina Sentencing and Policy Advisory Commission's Structured Sentencing Training and Reference Manual.

Figure A2: Misdemeanor Punishment Classes

| CLASS | PRIOR CONVICTION LEVEL | | |
|---|---|---|---|
| | **I** <br> No Prior Convictions | **II** <br> One to Four Prior Convictions | **III** <br> Five or More Prior Convictions |
| A1 | C/I/A <br> 1 - 60 days | C/I/A <br> 1 - 75 days | C/I/A <br> 1 - 150 days |
| 1 | C <br> 1 - 45 days | C/I/A <br> 1 - 45 days | C/I/A <br> 1 - 120 days |
| 2 | C <br> 1 - 30 days | C/I <br> 1 - 45 days | C/I/A <br> 1 - 60 days |
| 3 | C <br> 1 - 10 days | C/I <br> 1 - 15 days | C/I/A <br> 1 - 20 days |

Note: This chart shows how the class of the current offense and the number and severity of prior convictions determine different levels of punishment and sentence lengths. A = Active, I = Intermediate, C = Community. All misdemeanors classes offer options for non-active punishments. Source: North Carolina Sentencing and Policy Advisory Commission's Structured Sentencing Training and Reference Manual.

Figure A3: District and Superior Court Boundaries in 2009

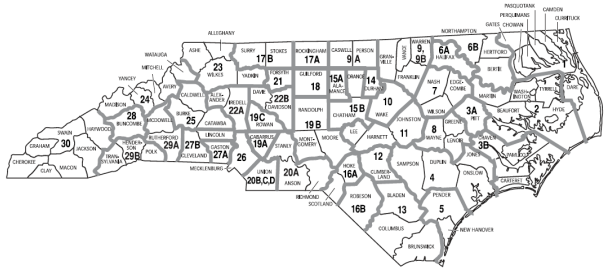 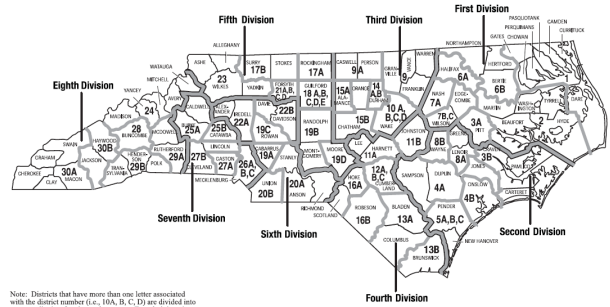

Note: These maps depict the geographic makeup of the two North Carolina court systems. The District Court (left) consists of 43 districts in which judges are randomly assigned to individual cases. The Superior Court (right) consists of 8 circuits (labeled "divisions" in this map), within which judges are rotated every 6 months to a new district. Source: North Carolina Courts Statistical and Operational Summary 2008-09.

Figure A4: The Effect of Mental Health Treatment on Recidivism over Time in the Balanced Sample

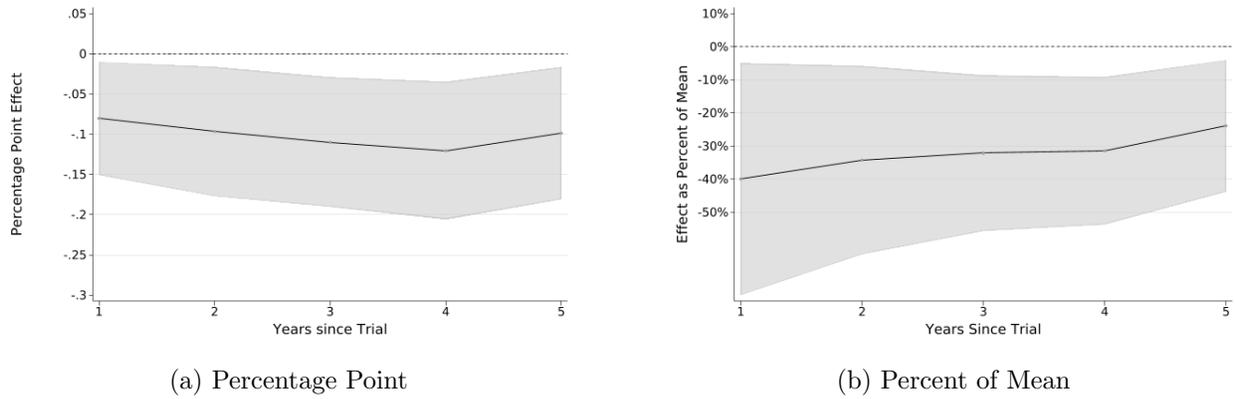

(a) Percentage Point
(b) Percent of Mean

Note: This figure plots the percentage point (left) and the percent (right) decline in recidivism due to mental health treatment, estimated from the main IV regression with full controls. Recidivism is defined first looking only within one year out from the trial, then between one and two years, then between two and three years, and so on up to five years. The estimation is carried out in a balanced sample that restricts to observations in the years 2004 and earlier. All specifications condition on the judge's tendency toward substance use disorder treatment as well as on court-time effects. The included controls are race, ethnicity, sex, age cubic, first-time offender, region of the United States, shift, day of week, month, year, attorney type, offense types, whether the offender had a prior offense in the past year, and county-level descriptors. Source: 1994-2009 North Carolina ACIS data.

Figure A5: The Effect of Mental Health Treatment on Recidivism over Time, OLS Estimates

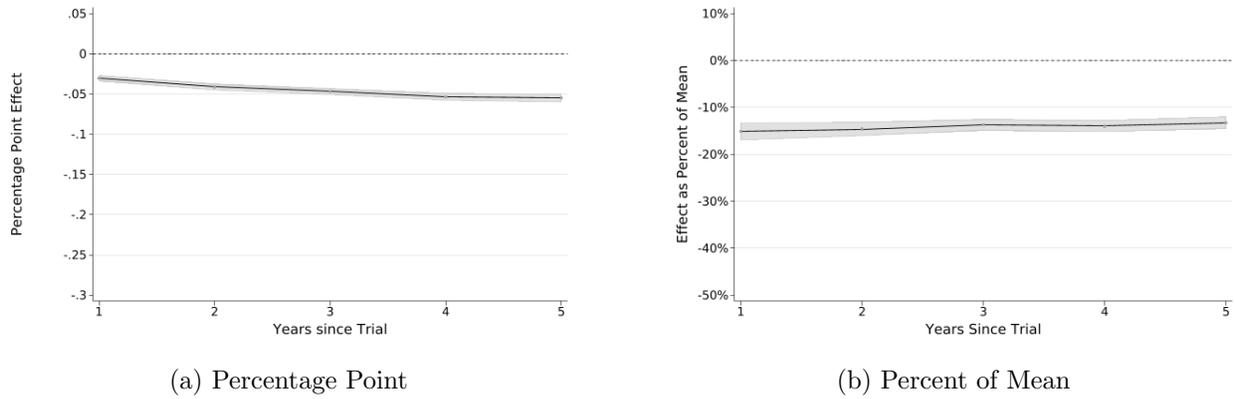

(a) Percentage Point

(b) Percent of Mean

Note: This figure plots the percentage point (left) and the percent (right) decline in recidivism due to mental health treatment, estimated from the OLS regression with full controls. Recidivism is defined first looking only within one year out from the trial, then between one and two years, then between two and three years, and so on up to five years. All specifications condition on the judge's tendency toward substance use disorder treatment as well as on court-time effects. The included controls are race, ethnicity, sex, age cubic, first-time offender, region of the United States, shift, day of week, month, year, attorney type, offense types, whether the offender had a prior offense in the past year, and county-level descriptors. Source: 1994-2009 North Carolina ACIS data.

# B  Appendix: Cost Benefit Calculations

This appendix details the calculations and sources that make up the cost-benefit estimates discussed in Section 6. Because there are a plethora of sources on various costs of crime, and each source approaches the accounting differently, I endeavor to create high and low estimates whenever possible, and report the average of those estimates in Section 6. As discussed in that section, I also endeavor to use costs specific to type of crime whenever available. I convert all monetary values to 2022 dollars using the BLS CPI inflation calculator.[20]

## B.1  Costs of Mental Health Treatment

The main difficulty in estimating the costs of the policy comes from the large variation in the possible components of treatment. For this analysis I define treatment as the combination of an initial evaluation, weekly meetings with a therapist for the duration of the probation sentence, and a 50% chance of taking antidepressants. For the low estimate I use the average community sentence length (4.8 months), the lower bound cost of a monthly supply of antidepressants ($4.60, Cherney (n.d.)), and the facility cost of therapy and evaluation ($80.17 and $127.12, DHHS (2013)). For the high estimate I use the average intermediate sentence length (9.6 months), the upper bound cost of a monthly supply of antidepressants ($149.50, Cherney (n.d.)), and the non-facility cost of therapy and evaluation ($87.53 and $161.75, DHHS (2013)).[21] In both cases, I adjust the cost of medication based on the Medicaid rebate, estimated at 0.231 by the Congressional Budget Office (CBO, 2022). The combination of initial evaluation and weekly therapy and medication for the duration of the sentence results in a baseline estimate within [$1,846.92, $4,237.64].

I make two adjustments to that baseline estimate of the costs. First, I address the likely individuals who are currently using Medicaid's services. In North Carolina, which did not expand Medicaid with the Affordable Care Act, eligibility for coverage comes predominantly from disability or responsibility for minors paired with an income requirement DHHS (2016). Using estimates from Glaze and Maruschak (2008) for the number of offenders who are parents of and financially responsible for children and were employed before prison, along with the distribution of earnings among employed offenders before prison from Garin et al. (2023), I calculate that %17 to %23 of probationers would be eligible for Medicaid due to being parents. Using estimates from Wang (2022) for the number of disabled prisoners and BLS (2014) for the employment rate among people with a disability, along with the distribution in earnings from Garin et al. (2023), I calculate that %38 of probationers would be eligible for Medicaid due to disability. I combine these and scale by the take-up rate of %61 among eligibles from Sommers et al. (2012) resulting in an estimated %34 to %38 of probationers who would have already been using Medicaid for help with medical costs. Subtracting the costs of providing medication and eight sessions of therapy (the maximum covered in a year by Medicaid) for those already-covered probationers, I arrive at my main estimate of the costs of treatment, $2,705.65 (interval: [$1,641.20, $3,770.09]).

The second adjustment addresses the potential behavioral response of taxpayers if the policy were paid for by an increase in taxes. I adjust the cost estimate by adding the deadweight loss due to funding the policy by increasing federal income taxes. I use Saez, Slemrod, and Giertz's estimate of the marginal excess burden per tax dollar at $0.195 (2012). Combining the effect of deadweight loss through taxation and the treatment that would have already been covered for about a third of probationers yields a final estimate of the costs of mental health treatment to the government of $3,233.25 per probationer (interval: [$1,961.24, $4,505.25]).

## B.2  Benefits of Mental Health Treatment

The benefits of mental health treatment among probationers likely include more than just the reduction in costs associated with lower crime. However, this paper's focus is on the effects of treatment on future criminal involvement, and lacks information on other potential benefits. I therefore proceed by quantifying only the decrease in costs associated with the decrease in crime, and consider these estimates to be a lower bound of the true monetary valuation of the benefits of mental health treatment.

---

[20] https://data.bls.gov/cgi-bin/cpicalc.pl

[21] Community and intermediate are the two types of probation sentences in North Carolina; intermediate involves an assigned probation officer, while community can be less regimented.

Even limiting to the reduction in costs associated with lower crime, quantifying the costs proves a daunting task. While there have been several attempts to assign monetary values to the criminal justice system, different methods of valuation can produce strikingly different cost estimates. In addition, the costs vary enormously by the type of crime and the punishment. For example, the direct operating costs associated with a year of prison average $42,862 in 2022 dollars, whereas the operating costs for a probationer average only $3,973 (Hunt, Anderson, and Saunders, 2017). Similar variation exists in the social costs associated with different crimes. With any crime that involves valuing a human life, estimates of the social cost are over $8.6 million; in contrast, selling or possessing stolen property has an average social cost of $686 per crime, and other crimes might have zero social cost. In order to incorporate that variation, I use cost estimates that are specific to crime types whenever possible. I also synthesize the existing literature into high and low estimates to address the sometimes large differences in estimates for ostensibly the same category of costs.

I consider four categories of costs associated with crime. The first is judicial costs, which include the operating costs of police forces, courts, jails, prisons, and probation and parole. The second is loss of tax revenue to the government due to prisoners not being employed. The third is impacts on offenders, such as through loss of earnings, court fines, and increased health issues due to incarceration. The fourth is the social costs, which are also called the costs of victimization or the intangible costs. These include medical bills and destroyed property of victims, as well as psychological effects due to fear of crime.

When sources break costs apart by crime, they often do so only for the FBI Part I crimes. Those include homicide, rape, robbery, aggravated assault, burglary, larceny, motor vehicle theft, and arson. Some sources provide partial estimates of total costs to more aggregated groups of Part II crimes, which are all crimes outside of the Part I crimes. To estimate costs for Part II crimes, I use the minimum cost from those partial estimates as a lower bound estimate and the average cost without disaggregating by crime type as the upper bound estimate. I assign costs to all Part I and Part II crime types, then aggregate those costs to the five offense groups used throughout this analysis (violent and property, financial and fraud, traffic and public order, drugs and alcohol, and miscellaneous). The "cost to offenders" category is not available disaggregated into crime types. Because most of the costs to offender are associated with incarceration, I weight the low and high estimates by the likelihood of incarceration within each offense group to form crime-type-specific estimates. Table A11 describes the sources used for each category of costs.

Table A11: Sources for Costs of Crime

| Type of Cost | Source |
| --- | --- |
| Judicial Costs | Hunt, Anderson, and Saunders (2017), McCollister, French, and Fang (2010) |
| Loss of Tax Revenue | McCollister, French, and Fang (2010) |
| Costs to Offenders | Binswanger et al. (2007), Garin et al. (2023), McLaughlin and Stokes (2002), McLaughlin et al. (2016), NIMH (2023), Rafael (2023), The Annie E. Casey Foundation (2023) |
| Social Costs | Heaton (2010), Hunt, Anderson, and Saunders (2017), McCollister, French, and Fang (2010) |

Note: This table lists the sources for various costs associated with crime, used to quantify the benefits from the reduction in crime due to mental health treatment.

The judicial, social, and lost tax revenue costs come directly from the listed sources, but the costs to offenders involve some additional calculations. The base estimates come from McLaughlin et al. (2016) and include lost wages while incarcerated, sexual abuse in prison, suicide risk, mortality risk, homelessness, moving, eviction, and adverse health. I also include reduced lifetime earnings from Garin et al. (2023) and conviction-related fines from Rafael (2023). Costs associated with the risks of homelessness, moving, eviction, and adverse health are given as annual costs over all offenders in the criminal justice system; I divide by the number of prisoners, probationers, and parolees in 2009 (5.8 million) to form a per-offense cost. The suicide and mortality costs are determined by comparing rates in prison to rates outside of prison; I adjust this by comparing to rates outside of prison among individuals with similar demographic characteristics. I use the suicide rate of Black individuals (13 out of 100,000) from NIMH (2023). I use the mortality rates from Binswanger et al. (2007) but adjust the comparison group's rate using the association between decreases in income and increases in mortality from McLaughlin and Stokes (2002), the median household income for Black individuals from The Annie E. Casey Foundation (2023), and the average income (conditional on employment) before incarceration from Garin et al. (2023). Because these costs to offenders are based on incarceration, I adjust the estimates by the likelihood of a prison sentence within each crime type.

Table A12: Benefits from Reduction in Crime

| Offense Type | Fraction of Sample | Judicial Costs | Costs to Offenders | Costs to Society | Loss of Revenue |
|---|---|---|---|---|---|
| Violent/Property | .28 | [$1,181, $14,497] | [$37,720, $77,583] | [$113,632, $217,089] | [$4,204, $7,235] |
| Financial/Fraud | .24 | [$4,227, $5,162] | [$10,764, $22,140] | [$1,382, $35,617] | [$627, $7,232] |
| Drugs/Alcohol | .22 | [$1,719, $4,910] | [$20,486, $42,137] | [$17,444, $86,247] | [$877, $11,836] |
| Traffic/Public Order | .25 | [$8,649, $10,383] | [$16,782, $34,517] | [$3,341, $54,766] | [$492, $10,423] |
| Miscellaneous | .01 | [$14, $14,741] | [$39,587, $81,424] | [$686, $91,543] | [$228, $18,301] |

Note: This table presents the lower and upper bound estimates for the costs of each crime, broken apart by four categories of costs and reported separately for five offense types. The costs are used to quantify the benefits from the reduction in crime due to mental health treatment. The sources for the costs are listed in Table A11.

To form the final estimate of the benefits of mental health treatment, I combine the costs of each crime type with the decrease in that crime type associated with treatment, then aggregate the resulting estimates based on the relative prevalence of each future crime type in the sample. Table A12 lists the low and high estimates for each crime in each category of costs. The variation in estimates comes from the different methods used by the various studies I rely on for cost information. Incorporating the estimated effect of mental health treatment on recidivism adds an additional source of uncertainty via the standard errors on the point estimates. Using the low estimates of crime costs, I estimate total benefits from the reduction in crime due to mental health treatment of about $9,185 (confidence interval: [$4,390, $13,985]). Using the high estimates of crime costs, I estimate the total benefits at about $23,077 (confidence interval: [$10,690, $35,484]).

# C  Appendix: Data Preparation

## C.1  ACIS

Data from the North Carolina Administrative Office of the Courts (AOC)'s Automated Criminal Infractions System (ACIS) is provided in a series of text files. I link those files using the case and charge keys. The result is a case-charge level data set. There are several entries per case for two reasons. First, one arrest may lead to several charges which are addressed in the same trial date. Second, the data include several entries per charge in order to describe additional sentence details. I collapse the data to the case level. I define most characteristics of the case using the most severe offense from among the charges. For example, the punishment (prison versus probation) is determined from the most severe charge. However, for special conditions of probation and certain aspects of the case, I use information from all charges. For example, a defendant received a mental health treatment mandate if any of the charges associated with the case involve a mental health treatment mandate. Similarly, a case is characterized as a domestic violence case if any of the component charges are related to domestic violence. When the same trial results in different disposition dates, I base my analysis on the earliest disposition date and associated judge. Later dates could represent future judgements based on the same case, such as probation violations.

I assign points following the Structured Sentencing Training and Reference Manual. Offenders are given more points if they have more prior sentences or if they are charged with a more severe offense. These points determine the class of sentence that offenders could receive – either active, intermediate, or community. Probation sentences are either community or intermediate, whereas prison sentences are active. Appendix Figures A1 and A2 depict how the points map to sentences for felonies and misdemeanors. Among felonies, offense class I (the lowest) with 0, 1-4, or 5-8 prior points result in C or I punishments. Misdemeanors of classes 1 and 2 with prior point levels 0 and 1-4, and class 3 with 0 prior points, all result in C or I punishments.

Identification of mental health treatment comes from the additional sentence details. I define offenders as being treated if any of the phrases "mental," "mntl," "mntal," "eval," "exam," "exm," "asses," "couns," "cnsl," "therap," "trt," "trea," "psy," "behavioral," "trmnt," "prescribed medicine," "prescribed meds," "psd meds," "mental health med," "mental med," "depress," "anger," "stress," "anxi," or "mood disord" appear in the detail field. I identify substance use disorder (SUD) treatment from the same additional sentence details, characterized by the phrases "drug trt," "drg trt," "drug trea," "drg trea," "drug eval," "drg eval," "alcohol," "DART," "subs," "sub abus," or "TASC." In both cases, I do not characterize the offender as being treated if the phrase "court does not recommend" also appears in the sentence field.

As my main definition of treatment, I make the additional restriction that phrases qualified with 'substance abuse' or 'drug' - for example, if the phrase 'treatment' appears but is preceded by 'drug' - do not contribute toward mental health treatment cases. Specifically, I remove cases from the treatment definition if I have also categorized them as SUD treatment *and* they do not include any of "mental," "mntl," "mntal," "couns," "therap," "psy," "behavioral," "mental health med," "mental med," "depress," "anxi," and "mood disord."

I consider several alternate definitions of treatment, including one based only on the above phrases without removing likely SUD-only cases. I broaden the definition further by including offenders likely in some sort of program, characterized by "program," "medical issu," "medical eval," "medical prob," "medical trea," "residential," "inpatient," or "rehab." This represents the most broad definition of treatment. I also narrow it in several ways. From the baseline measure, I remove all cases that are also categorized as SUD treatment. In another more narrow definition, I remove cases that might have been sent to a mental health court, characterized by "S.T.E.P.," "mental health court," "by mhc," "community resource court," "to crc," "in crc," "crc court," "crc prog," "complete crc," "by crc," "completed crc," or "attend crc." These represent the most narrow definition.

I create longitudinal links using name and birth date in order to track offenders' cases and identify recidivism. The data contain inconsistencies in the spelling of names. I correct errors in which similar names have the same other key characteristics. For example, I treat "Vasquez, Jose Perrero Arturo" and "Vasquez, Jose Perrero" as the same person if they have the same birth date, race, and sex. Similarly, I correct birth date inconsistencies when other key characteristics are the same. For example, I treat "011754" and "101754" as the same person when the name, race, and zip code exactly match. When birth dates disagree as in this example, I calculate the defendant's age using the most frequent birth date among that defendant's cases.

## C.2 Judge Schedules and District Maps

I create a list of judges and their tenures using yearly documents from the NC Judicial Department website.[22] The same website provides judicial district maps, which I use to determine the superior court circuits and for verification of judge districts. Spring and Fall schedules are also available, from which I draw the exact judges who were presiding in a given court at a given time of day and day of week.

I combine the supplemental information with the ACIS in several stages. First, I add information about the District and Superior districts and the Superior circuit based on the county and year associated with the trial. Next, I add information about the schedules of the judges based on the judge, district, circuit, court, year, and season associated with the case. When doing so, I correct obvious spelling errors in the ACIS judge initials, such as cases where the "O" was typed as a "0." In addition, the judge is only identified by a two- or three-letter initial in the ACIS, whereas my supplemental information gives the full name of the judge. I create both two- and three-letter versions of the judge initials from the full name in order to combine files. When the ACIS is missing information about court, circuit, and district, I instead add the judge schedule information using only judge, year, and season.

## C.3 Sample Restrictions

Table A13: Sample Reductions

|  | Observations | % Dropped | % Remaining of Original |
| --- | --- | --- | --- |
| Start | 2108124 | 0.0 | 100.0 |
| Key variables | 1612161 | 23.5 | 76.5 |
| Adults | 1572485 | 2.5 | 74.6 |
| Struct. Sent. Period | 1440145 | 8.4 | 68.3 |
| No Drug Court | 1439406 | 0.1 | 68.3 |
| Judge Cases | 1375395 | 4.4 | 65.2 |
| Probation Charge | 1039994 | 24.4 | 49.3 |
| Enough Cases per Judge | 1027754 | 1.2 | 48.8 |

Note: This table walks through each sample restriction, starting from the raw data and ending with a sample of individuals who were only at risk of a probation sentence (no chance of a prison sentence) due to the Structured Sentencing Laws in place from October 1, 1994 to December 1, 2009.

Table A13 displays the sample restrictions made. After collapsing to the case level, there are 2,108,141 offender-trial observations in the raw data.[23] I limit the sample to offenders with key information, including an individual identifier, a judge, and the judge's district or court. I require offenders in my sample to be adults, which in North Carolina means 16 or older, and remove individuals with unrealistic ages (older than 100). I limit the time period to criminal court cases from October 1, 1994 to December 1, 2009, during the first era of structured sentencing. I remove those in "A" and "I/A" punishment classes based on their charged offense, losing 20 percent compared to the step before. This leaves only individuals whose only available sentence is probation. I also remove judges with fewer than 10 cases in a year in order to have enough variation to construct the instrument. After implementing these restrictions, the sample includes 1,028,087 cases, 51,404 (5 percent) of which involved a sentence to seek mental counseling.

---

[22]https://www.nccourts.gov/documents/publications
[23]The original sample included several entries per case in order to describe additional sentence details.